\documentclass[useAMS,usenatbib]{mn2e}
\usepackage{amsmath}
\usepackage{natbib}
\usepackage{subfig}
\usepackage{graphicx}

\title[Dust trapping by vortices in self-gravitating discs]{Planetesimal formation
in self-gravitating discs -- dust trapping by vortices}
\author[Gibbons et al.]{P. G. Gibbons$^{1}$\thanks{E-mail: pgg@roe.ac.uk}, G. R. Mamatsashvili$^{2,3}$ and W. K. M. Rice$^{1}$\\
$^{1}$SUPA, Institute for Astronomy, Royal Observatory, Blackford Hill, Edinburgh EH9 3HJ\\
$^{2}$Department of Physics, Faculty of Exact and Natural Sciences,
Tbilisi State University, Tbilisi 0179, Georgia\\
$^{3}$Abastumani Astrophysical Observatory, Ilia State University,
Tbilisi 0162, Georgia}
\begin{document}
\date{Accepted ????. Received ????? ; in original form ?????}
\pagerange{\pageref{firstpage}--\pageref{lastpage}} \pubyear{2002}
\maketitle
\label{firstpage}

\begin{abstract}
\noindent The mechanism through which meter-sized boulders grow to
km-sized planetesimals in protoplanetary discs is a subject of
active research, since it is critical for planet formation. To avoid
spiralling into the protostar due to aerodynamic drag, objects must
rapidly grow from cm-sized pebbles, which are tightly coupled to the
gas, to large boulders of 1-100m in diameter. It is already well
known that over-densities in the gaseous component of the disc
provide potential sites for the collection of solids, and that
significant density structures in the gaseous component of the disc
(e.g., spiral density waves) can trap solids efficiently enough for
the solid component of the disc to undergo further gravitational
collapse due to their own self-gravity. In this work, we employ the
{\scriptsize PENCIL CODE} to conduct local shearing sheet
simulations of massive self-gravitating protoplanetary discs, to
study the effect of anticyclonic transient vortices, or eddies, on
the evolution of solids in these discs. We find that these types of
structures are extremely efficient at concentrating small and
intermediate-sized dust particles with friction times comparable to,
or less than, the local orbital period of the disc. This can lead to
significant over-densities in the solid component of the disc, with
density enhancements comparable to, and even higher, than those
within spiral density waves; increasing the rate of gravitational
collapse of solids into bound structures.
\end{abstract}


\section{Introduction}

Large-scale spiral density waves commonly arise in a protoplanetary
disc due to the gravitational instabilities at the early stage of
disc life, when it is sufficiently massive. It has been known for
some time that the density waves can act to transport angular
momentum outwards, allowing mass to accrete onto the protostar, and
could well be the primary transport mechanism provided thus by disc
self-gravity during the earliest stages of star formation
\citep[e.g.,][]{Lin1987,Rice2010}.

The other perturbation type in discs that has received much
attention, partly in connection with planetesimal formation, are
vortices. They can arise in both self-gravitating and
non-self-gravitating discs by several means. A localized radial
structure in the disc can trigger the Rossby wave instability, which
in turn results in the formation of vortices
\citep[e.g.,][]{Lovelace1999,Li2000,Lyra2009,Meheut2010,Meheut2012a,Lin2012a,Regaly2012}.
Vortices can also be generated as a result of other hydrodynamic
instabilities, such as the baroclinic instability
\citep{Klahr2003,Petersen2007,Lesur2010, Raettig2013} or the
Kelvin-Helmholtz instability of seed vorticity strips
\citep{Lithwick2007}. Although the dynamics and evolution of
vortices in non-self-gravitating discs has been under active
investigation for a while, since the first idea of their relevance
to planet formation by \cite{Barge1995}, systematic studies of
vortex dynamics in self-gravitating discs are quite recent
\citep{Mamatsashvili2009,Lyra2009,Lin2012b,Lovelace2013,Ataiee2014}.
It was commonly thought that spiral density waves are the only
perturbation type present in self-gravitating discs, and the
existence of other modes also participating in the overall disc
dynamics had been ignored in many studies of self-gravitating discs.

Developments in the theory of non-self-gravitating discs, however,
showed that vortices can be (linearly) coupled with, and excite,
spiral density waves due to the disc's differential rotation
\citep{Bodo2005, Johnson2005, Heinemann2009}. This effect was
demonstrated to be even more efficient in the presence of
self-gravity \citep{Mamatsashvili2007}. Moreover, a linear analysis
by \cite{Mamatsashvili2007} showed that in fact vortical mode is
also subject to the influence of self-gravity and can exhibit
gravitational instability with amplification factors comparable to
that of density waves. This clearly implies that vortices are as
important as density waves in forming a complete dynamical picture
of self-gravitating discs.

Subsequent nonlinear dynamics and evolution of vortices varies
greatly in non-self-gravitating and self-gravitating systems. In
non-self-gravitating discs, small-scale (anticyclonic) vortices,
soon after formation, smoothly merge into each other and increase in
size \citep[e.g.,][]{Li2001,Godon1999,Godon2000,
Klahr2003,Umurhan2004,Johnson2005,Shen2006}. When the vortices reach
a few scale height, compressibility becomes important, causing them
to emit spiral density waves - due to the above-mentioned mode
coupling phenomenon - and, consequently, slowly decay on the
time-scale of several hundred orbital periods
\citep{Johnson2005,Shen2006,Bodo2007}. In spite of this shock
dissipation, anticyclonic vortices can be considered long-lived,
coherent structures in discs. However, they can be subject to radial
migration \citep{Paardekooper2010} and/or to the elliptical
instability \citep{Lesur2009, Richard2013}, the typical time-scale
of which is larger than vortex turnover/orbital time, as well as to
destruction by the magnetorotational instability in magnetized
regions \citep{Lyra2011}.

The vortex merging is resisted under the influence of disc
self-gravity and smaller size vortices are favoured
\citep{Mamatsashvili2009,Lyra2009,Lin2012b}. In some sense,
self-gravity acts to oppose the inverse cascade of spectral energy
to larger scales, as occurs in the non-self-gravitating case, and,
instead, scatters it to smaller scales. The local shearing sheet
simulations of radially unstructured (i.e., in the absence of
baroclinic or Rossby wave instabilities) discs by
\cite{Mamatsashvili2009} demonstrate that vortices in a quasi-steady
gravitoturbulence are transient, short-lived structures undergoing
repeating cycles of formation, growth to sizes comparable to the
local Jeans length, and eventual shearing and destruction due to the
effects of self-gravity and Keplerian shear. This process lasts a
few orbital periods, and results in a very different, compared to
the non-self-gravitating case, evolutionary picture, with many
small, less organized vortices (eddies) in various stages of
evolution, rather than the relatively larger scale well organized
vortices gradually growing via slow mergers. On the other hand, in
global simulations by \cite{Lyra2009,Lin2012b} discs are radially
structured and, although they remain laminar (without
gravitoturbulence), develop vortices via Rossby wave instability at
vortensity minimum. Due to self-gravity, these vortices have a
smaller azimuthal extent (i.e., higher azimuthal wavenumbers)
compared with similar ones without self-gravity. What is noteworthy,
however, vortices in both local and global simulations share a
common feature -- they produce over-densities at their centres
coinciding with the minima of Toomre's parameter $Q$. These
over-densities are imposed on the density variations due to shocks
of density waves emitted by these vortices.

In massive discs, the gravitational instabilities have been shown to
cause the disc to fragment, possibly leading to the direct formation
of gas giant planets
\citep{Boss1998,Gammie2001,Rice2003,Paardekooper2012}. Although a
massive protoplanetary disc is thought to be present at early times
in the star formation process, it is not clear that such discs can
cool fast enough due to radiative transfer to directly form giant
planets via this mechanism
\citep{Rafikov2005,Boley2006,Clarke2009,Rice2009}. However, even if
the effective cooling times present within discs are too long to
allow giant planets to form via fragmentation, as mentioned above,
protoplanetary discs are very likely self-gravitating in their early
stages \citep{Lin1987, Lin1990}, sometimes even at the T Tauri stage
\citep{Eisner2005, Andrews2007}. In this case, indirect means of
planet formation -- accumulation of dust particles in gaseous
density structures that arise as a result of the gravitational
instability -- could be at work. As outlined above, such structures
can be either due to density waves or to smaller scale vortices.

In low mass discs, where self-gravity is negligible, vortices have
been shown to have a significant influence on the local evolution of
dust particles \citep[e.g.,][]{Johansen2004,Klahr2006, Meheut2012b,
Lyra2013, Fu2014,Zhu2014}. In these simulations, not including
self-gravity, dust grains with a wide range of sizes were shown to
be trapped within the centres of the vortex structures with the
strongest concentration occurring for particles with
friction/stopping times comparable to the local orbital period. In
other words, a smooth, sufficiently long-lived vortex is indeed able
to effectively trap dust particles in its core, possibly
accelerating planetesimal formation.

To date, however, the evolution of dust particles in
self-gravitating discs has been primarily studied in a large scale
setting, concentrating on the effect spiral density waves, in the
gaseous component of the disc, have on dust particles.
\citet{Rice2004,Rice2006} used global simulations to show that the
presence of spiral density waves in the disc can have a dramatic
effect on the particle layer. In \citet{Rice2004}, the dust
particles are assumed to interact with the disc solely via an
aerodynamic drag force. The particles, especially those
with stopping times comparable to the local orbital period, are
shown to become trapped within the density wave structures that form
in the disc due to self-gravity. This causes the local density of solid
material within the density waves to rise by an order of magnitude
or more compared to the average. The gas pressure gradient changes from
positive to negative across the density wave structure, creating
sub-Keplerian velocities on one side of the wave, and
super-Keplerian on the other. As a result, the drag force causes dust
grains to drift toward the density/pressure maxima at the crest of
density waves.

\citet{Rice2006} generalized these findings by including the
gravitational effects of the solid particles themselves, which allow
these particle enhancements to become even more pronounced,
resulting in gravitationally bound clumps, or planetesimals, forming
in the disc. More recently, these results were expanded on in
\cite{Gibbons2012,Gibbons2014} using higher resolution
two-dimensional (2D) local shearing sheet simulations of
self-gravitating gaseous discs, treated as in related local studies
\citep{Gammie2001,Rice2011,Paardekooper2012}, with added numerical
super-particles to model dust-particle dynamics. These studies found
that the spiral density waves are very efficient at trapping
relatively small, from several centimeter up to meter-sized
particles, whose stopping times are comparable to the local orbital
period. Very tightly bound clumps of particles can form in the
crests of these spiral waves due to the strong gravitational
attraction between particles at such high concentrations. These
processes can result in a substantial population of large,
planetesimal-sized solid objects being rapidly formed in the early
stages of disc evolution, paving the way for terrestrial and gas
giant planet formation via the traditional core-accretion model
\citep[e.g.,][]{Pollack1996}.

Dust particles can also become trapped by pressure maxima associated
with (anticyclonic) vortices and, as mentioned above, this process
has been extensively studied in non-self-gravitating discs. The role
of vortices in trapping dust in self-gravitating discs is less
understood and needs further investigation, because of the different
character of vortex evolution with self-gravity. Particle
concentration by vortices, as well as vortex dynamics itself in the
presence of self-gravity, was first addressed by \cite{Lyra2009} via
global disc simulations. However, in their set-up vortices are
generated by the Rossby wave instability in an otherwise laminar
disc and have a long-lived regular nature - as in the
non-self-gravitating case - though smaller azimuthal size due to the
effect of self-gravity. Here, we consider a different situation -- a
self-gravitating disc in a quasi-steady gravitoturbulent state,
without continuous driving by global baroclinic or Rossby wave
instabilities, and analyse the behaviour of embedded dust particles
using the shearing sheet approximation. We would like to note that
on shorter (comparable to the disc scale height or less)
length-scales just such a gravitoturbulent state, rather than a
regular/coherent one, is more common in self-gravitating discs
\citep[e.g.,][]{Gammie2001,Durisen2007}. As shown by
\citet{Mamatsashvili2009}, this state consists of transient
short-lived small-scale vortices producing over-dense regions
overlaid on density wave structures. In other words, we extend this
study by including dust particles with a goal of understanding the
role these vortical structures play in dust evolution. Dust dynamics
in the presence of only density waves in the shearing sheet has been
explored in our previous papers \citep{Gibbons2012,Gibbons2014},
where larger domain sizes were used. Here, we take a smaller size domain
that enables us to zoom in on smaller scale vortical structures
and examine their dust trapping capability against the backdrop of
density variations due to the density waves (shocks). This will help
us to establish the role of vortices in planet formation process in
self-gravitating discs.

The paper is organized as follows. In Section 2, we outline the
physical set-up and initial conditions adopted in our simulations.
In Section 3, we discuss the evolution of the dust particles in the
the vortex structures in the gas. Finally, summary and discussions
are given in Section 4.

\section[]{Model}

\label{Model} To investigate the dynamics of solid particles
embedded in a self-gravitating protoplanetary disc, we solve the 2D
local shearing sheet equations for gas on a fixed grid, including
disc self-gravity, together with the equations of motion of solid
particles coupled to the gas through aerodynamic drag force. We also
include the self-gravity of particles necessary to examine their
collapse properties. As a main numerical tool, we employ the
{\scriptsize{PENCIL CODE}} \footnote{See
http://code.google.com/p/pencil-code/}. It is a sixth order spatial
and third order temporal finite difference code (see
\citet{Brandenburg2003} for full details). The {\scriptsize{PENCIL
CODE}} treats solids as numerical super-particles
\citep{Johansen2006,Johansen2011}.

In the shearing sheet approximation, disc dynamics is studied in the
local Cartesian coordinate frame centred at some fiducial radius,
$r_0$, from the central object, and rotating with the disc's angular
frequency, $\Omega$, at this radius. In this frame with the $x$-and
$y-$axis pointing, respectively, in the radial and azimuthal
directions, the disc's differential rotation manifests itself as an
azimuthal parallel flow characterized by a linear shear, $q$, of
background velocity along the $x-$axis, ${\bf u}_0=(0,-q\Omega x)$.
The equilibrium surface densities of gas, $\Sigma_0$, and particles,
$\Sigma_{p0}$, are spatially uniform. Since the disc is cool, and
therefore thin, the aspect ratio is small, $H/r_0\ll 1$, where
$H=c_s/\Omega$ is the disc scale height, and $c_s$ is the gas sound
speed. The local shearing sheet model is based on the expansion of
the basic 2D hydrodynamic equations of motion to the lowest order in
this small parameter, assuming that the disc is also razor thin
\citep[see e.g.,][]{Gammie2001}.

Our simulation domain spans the region $-L_x/2 \leq x \leq L_x/2$,
$-L_y/2 \leq y \leq L_y/2$. As is customary, we adopt the standard
shear-periodic boundary conditions well approbated in shearing sheet
studies of discs \citep[e.g.,][]{Gammie2001,Johansen_Youdin2007}.
Namely, for any variable $f$, including azimuthal velocity with
background flow subtracted, we have
\[
f(x,y,t) = f(x+L_x,y-q\Omega L_xt,t),~~~~~~(x-{\rm boundary})
\]
\[
f(x,y,t) = f(x,y+L_y,t),~~~~~~~(y-{\rm boundary})
\]
The shear parameter is $q=1.5$ for the Keplerian rotation profile
adopted in this paper.

\subsection{Gas density}

In this local model, the continuity equation for the vertically
integrated gas density $\Sigma$ is
\begin{equation}
\frac{\partial\Sigma}{\partial t} + {\bf\nabla}\cdot(\Sigma{\bf u})
-q\Omega x\frac{\partial\Sigma}{\partial y}-f_D(\Sigma) = 0
\end{equation}
where ${\bf u}=(u_x,u_y)$ is the gas velocity relative to the
background Keplerian shear flow ${\bf u}_0$. Due to the high-order
numerical scheme of the {\scriptsize{PENCIL CODE}} it also includes
a diffusion term, $f_D$, to ensure numerical stability and capture
shocks,
\[
f_D = \zeta_D(\nabla^2 \Sigma +\nabla \textrm{ ln } \zeta_D \cdot
\nabla\Sigma).
\]
Here the quantity $\zeta_D$ is the shock diffusion coefficient
defined as
\[
\zeta_D = D_{sh} \langle \max_3[(-\nabla\cdot {\bf u})_+]
\rangle(\delta x)^2 \label{shock}
\]
where $D_{sh}$ is a constant, characterizing the strength of shock
diffusion as outlined in Appendix B of \citet{Lyra2008}, and $\delta
x=L_x/N_x$ is the grid cell size.

\subsection{Gas velocity}

The equation of motion for the gas relative to the unperturbed
Keplerian flow takes the form
\begin{multline}
\frac{\partial {\bf u}}{\partial t}+({\bf u}\cdot\nabla){\bf
u}-q\Omega x \frac{\partial {\bf u}}{\partial y} = -\frac{\nabla
P}{\Sigma} - 2\Omega {\bf\hat{z}}\times{\bf u}+q\Omega u_x
{\bf\hat{y}}\\-\frac{\Sigma_p}{\Sigma}\cdot\frac{{\bf u}-{\bf
v}_p}{\tau_f} - \nabla\psi +\bf f_\nu (u), \label{gvel1}
\end{multline}
where $P$ is the vertically integrated pressure, $\Sigma_p$ is the
surface density of particles, $\psi$ is the gravitational potential produced jointly by the gas
and particle surface densities (see equation 6). The left hand
side of equation (2) describes the advection by the velocity field,
{\bf u}, itself and by the mean Keplerian flow. On the right hand
side, the first term is the pressure force, the second and third
terms represent the Coriolis force and the effect of shear,
respectively, and the fourth term describes the back-reaction
exerted on the gas by the dust particles due to aerodynamic drag
force \citep[see e.g.,][]{Lyra2009,Johansen2011}. This force is proportional to
$\Sigma_p$ and to the difference between the velocity of particles ${\bf v}_p$ and
the gas velocity ${\bf u}$ and is inversely proportional to the
stopping, or friction time, $\tau_f$, of particles. The fifth term
represents the force due to self-gravity of the system. Finally, the
code includes an explicit viscosity term, $\bf f_\nu$,
\begin{align*}
{\bf f_\nu} =&  \nu(\nabla^2{\bf u} +
\frac{1}{3}\nabla\nabla\cdot{\bf u}
+ 2{\bf S}\cdot \nabla \textrm{ln }\Sigma) \nonumber \\
& + \zeta_\nu[\nabla(\nabla\cdot{\bf u})+ (\nabla \textrm{ln }\Sigma
+ \nabla\textrm{ln }\zeta_\nu)\nabla\cdot{\bf u}],
\end{align*}
which contains both shear viscosity and a bulk viscosity for
resolving shocks. Here {\bf S} is the traceless rate-of-strain
tensor
\[
S_{i j} = \frac{1}{2}\left(\frac{\partial u_i}{\partial
x_j}+\frac{\partial u_j}{\partial x_i} - \frac{2}{3}\delta_{i
j}\nabla\cdot{\bf u}\right)
\]
and $\zeta_{\nu}$ is the shock viscosity coefficient analogous to
the shock diffusion coefficient $\zeta_D$ defined above, but with
$D_{sh}$ replaced by $\nu_{sh}$.

\subsection{Entropy}

The {\scriptsize{PENCIL CODE}} uses entropy, $s$, as its main
thermodynamic variable, rather than internal energy, $U$.
The equation for entropy evolution is
\begin{equation}
\frac{\partial s}{\partial t}+({\bf u}\cdot\nabla)s - q\Omega
x\frac{\partial s}{\partial y} = \frac{1}{\Sigma
T}\left(2\Sigma\nu{\bf S}^2 - \frac{\Sigma
c_s^2}{\gamma(\gamma-1)t_c} + f_{\chi}(s)\right)
\end{equation}
where the first term on the right hand side is the viscous heating
term and the second term is an explicit cooling. Following
\cite{Gammie2001,Rice2011,Paardekooper2012}, we assume
the effective cooling time $t_c$ to be constant throughout the simulation
domain and take its value $t_c = 20\Omega^{-1}$, which is
sufficiently large for the disc to avoid fragmentation and settle into a
quasi-steady self-regulated state. The final term on the right hand
side, $f_{\chi}(s)$, is a shock dissipation term analogous to that
outlined for the density.

\subsection{Dust particles}

The dust particles are treated as a large number of numerical
super-particles \citep{Johansen2011} with positions ${\bf
r}_p=(x_p,y_p)$ on the grid and velocities ${\bf v}_p=({\rm
v}_{p,x},{\rm v}_{p,y})$ relative to the unperturbed Keplerian
rotation velocity, ${\bf v}_{p,0}=(0,-q\Omega x_p)$, of particles in
the local Cartesian frame. These are evolved according to
\begin{equation}
\frac{\mathrm{d}{\bf r}_p}{\mathrm{d}t} = {\bf v}_p - q\Omega
x_p{\bf \hat{y}}
\end{equation}
\begin{equation}
\frac{\mathrm{d}{{\bf v}_p}}{\mathrm{d}t} = - 2\Omega
{\bf\hat{z}}\times{\bf v}_p+q\Omega {\rm v}_{p,x}
{\bf\hat{y}}-\nabla\psi+\frac{{\bf u} - {\bf v}_p}{\tau_f}.
\label{parvel}
\end{equation}
The first two terms on the right hand side of equation (5) represent
the Coriolis force and the non-inertial force due to shear. The
third term is the force exerted on the particles due to the common
gravitational potential $\psi$. The fourth term describes the drag
force exerted by the gas on the particles which arises from the
velocity difference between the two. Unlike the gas, the particles
do not feel the pressure force. In the code, the drag force on the
particles from the gas is calculated by interpolating the gas
velocity field to the position of the particle, using the second
order spline interpolation outlined in Appendix A of
\citet{Youdin2007}. The back-reaction on the gas from particles in
equation (2) is calculated by the scheme outlined in
\citet{Johansen2011}.

\subsection{Self-gravity}

The gravitational potential in the dynamical equations (2) and (5)
is calculated by inverting Poisson equation for it, which contains
on the right hand side the gas plus particle surface densities in a
razor thin disc \citep[e.g.,][]{Lyra2009}
\begin{equation}
\Delta\psi = 4\pi
G(\Sigma-\Sigma_0+\Sigma_p-\Sigma_{p0})\delta(z)\label{Poisson}
\end{equation}
using the Fast Fourier Transform (FFT) method outlined in the
supplementary material of \citet{Johansen2007}. Note that the
perturbed gas, $\Sigma-\Sigma_0$, and particle,
$\Sigma_p-\Sigma_{p0}$, surface densities enter equation (6), since
only the gravitational potential associated with the perturbed
motion (and hence density perturbation) of both the gaseous and
solid components determine gravity force in equations (2) and (5).
Here, the surface density is Fourier transformed from the
$(x,y)$-plane to the $(k_x,k_y)$-plane without the intermediate
co-ordinate transformation performed by \citet{Gammie2001}. For this
purpose, a standard FFT method has been adapted to allow for the
fact that the radial wavenumber $k_x$ of each spatial Fourier
harmonic depends on time as $k_x(t) = k_x(0) + q\Omega k_yt$ in
order to satisfy the shearing sheet boundary conditions \citep[see
also][]{Mamatsashvili2009}.

\subsection{Units and Initial Conditions}

We normalise the quantities by setting $c_{s0}= \Omega=\Sigma_0= 1$.
The time and velocity units are $[t] = \Omega^{-1}$ and $[v] =
c_{s0}$, resulting in the orbital period, $T_{orb} = 2\pi$. The unit
of length is the scale height, $[l] = H = c_{s0}/\Omega$. The
initial Toomre parameter $Q=c_{s0}\Omega/\pi G \Sigma_0$ is equal to
1 throughout the domain. This sets the gravitational constant $G$ =
$\pi^{-1}$. The surface density of gas is initially set to unity
everywhere in the sheet. The simulation domain is a square with
dimensions $L_x = L_y=20G\Sigma_0/\Omega^2$ and is divided into a
grid of $N_x\times N_y = 1024\times 1024$ cells. This choice of
units sets the domain length $L_x = 20H/\pi Q \approx 6.37H$ and the
corresponding grid cell size $\delta x=\delta y=0.0062H$.

The gas velocity field is initially perturbed by some small random
fluctuations with the rms amplitude $ \sqrt{\langle{\bf u}^2\rangle}
= 10^{-3}$. We take the viscosity and diffusion coefficients to be
$\nu = 10^{-4}$ and $\nu_{sh} = D_{sh} = 10.0$, this ensures
numerical stability across the shock fronts, without washing them
out. We use $10^6$ particles, split evenly between five friction
times, $\tau_f = [0.01,0.1,1,10,100]\Omega^{-1}$. The dust particles
are initially randomly placed throughout the domain with zero
velocity and do not interact with the gas or other dust particles
for the first 10 orbital periods. At this point, once the gas has
settled into a quasi-steady state, which is independent of its
initial conditions, the drag forces between the dust particles and
gas are introduced. The system is then evolved for a further 30
orbits, before the particles' self-gravity is introduced. The
initial value for the dust-to-gas surface density ratio,
$\Sigma_{p0}/\Sigma_0$, is taken to be 0.01 for each particle
species.

\begin{figure}
\includegraphics[width = \columnwidth]{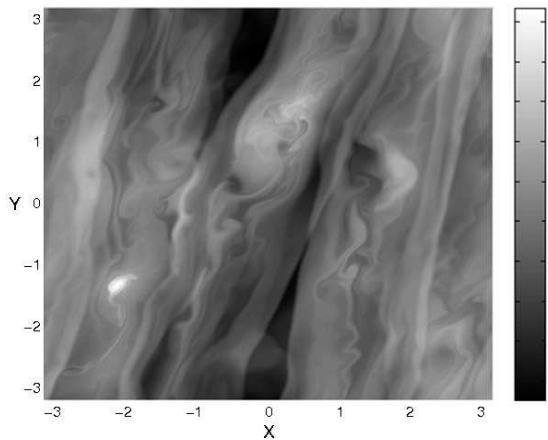}
\caption{Logarithmic surface density of the gas after 40 orbits,
just before the self-gravity of the particles is introduced. The
disc has already settled into a quasi-steady gravitoturbulent state.
Superimposed on the larger scale density waves are the smaller scale
over-dense structures due to irregular vortices (see Fig.
\ref{vor_field}). The distribution of dust particles at the same
time is shown in Fig. \ref{surf_density}.} \label{gas_dens}
\end{figure}
\begin{figure}
\includegraphics[width = \columnwidth]{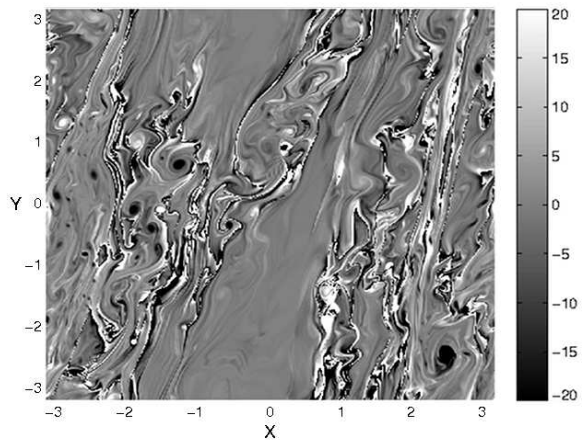}
\caption{PV field, $I$, in the quasi-steady state at the same time
as in Fig. \ref{gas_dens}. Numerous small-scale, irregularly shaped,
anticyclonic vortices (eddies), corresponding to negative PV regions
(black dots and curly diffuse areas) have developed in the sheet.
The over-dense structures produced by these vortices are clearly
seen in the gas density field shown in Fig. \ref{gas_dens}. The
colour-map is restricted to the range $-20\leq I\leq 20$ to better
emphasize/visualize the presence of the vortical structures.}
\label{vor_field}
\end{figure}

\section{Results}

\subsection{Gas Evolution}

The evolution of the gaseous component of the disc is in good
agreement with that observed in previous analogous studies of
self-gravitating discs in the shearing sheet with an imposed
constant cooling time. The initial perturbations grow and develop
into nonlinear fluctuations in velocity, surface density and
potential. Shocks then form, which proceed to heat the gas, while
the cooling acts to reduce the internal energy and entropy of the
gas. Density structures develop, which are sheared out by
differential rotation. Once the system has completed several orbits,
the heating generated by the shocks is balanced by the cooling term,
and the system settles into a quasi-steady, self-regulated state. In
this gravitoturbulent state, the thermal, kinetic and gravitational
energies of the disc are on average constant over time. The
saturated value of the shear stress parameter $\alpha$ is
proportional to the inverse cooling time, while the amplitude of the
density fluctuations to the square root of the inverse cooling time
\citep{Cossins2009,Rice2011}. Figure \ref{gas_dens} shows the
surface density of the gas after 40 orbits in the fully developed
quasi-steady gravitoturbulence. Here, the large-scale shocks due to
density waves are visible along with the smaller scale structures
associated with vortices.\footnote{In the Appendix, we show, by calculating the
autocorrelation function for the PV field in the quasi-steady state, that the shortest
correlation length for these vortices is more than 10 times larger than
the grid cell size, implying that they are sufficiently resolved in
our simulations.} The dynamics of these vortices in the
gravitoturbulent state was investigated by
\citet{Mamatsashvili2009}, which we recap below.

The vortical structures in the gas are characterized by the
potential vorticity (PV),
\[
I \equiv \frac{{(\bf \nabla \times u)\cdot\hat{z}}
+(2-q)\Omega}{\Sigma}.
\]
During the burst phase, initial small-scale positive and negative PV
regions are strongly sheared into strips, but negative PV
(anticyclonic) regions start to wrap up into vortex-like structures
due to the nonlinear Kelvin-Helmholtz instability
\citep{Lithwick2007}. The positive PV regions remain sheared into
strips, showing no signs of vortex formation during the entire
course of evolution. Only these anticyclonic regions, having
negative PV, are able to survive in the flow by taking the form of
vortices, though they are not as regular/coherent as those occurring
in non-self-gravitating discs
\citep[e.g.,][]{Umurhan2004,Johnson2005,Shen2006}. These smaller
scale, irregular anticyclonic vortices, or eddies, characterized by
negative PV, which have developed in the quasi-steady state are
clearly seen in the PV field in Fig. \ref{vor_field}. They induce
spatial variations (structures) with a similar form in the gas
surface density due to compressibility and self-gravity, as clearly
seen in Fig. \ref{gas_dens}. One can say that in the presence of
self-gravity, the PV field makes its ``imprint'' in the density
field. This is in contrast to the case with multiple nonlinearly
interacting vortices in the non-self-gravitating shearing sheet,
where only density waves (shocks) shed by them are seen in the gas
density field \citep{Shen2006}. Comparing Figs. \ref{gas_dens} and
\ref{vor_field}, we notice that those vortices with smaller, by
absolute value, PV give rise to higher over-densities than those
with larger, by absolute value, PV (for example, small white region
in the surface density map near the lower left corner). These
over-densities at some locations are even larger than the density
variations near the shock fronts.

\begin{figure*}
\subfloat{\includegraphics[trim = 10mm 5mm 10mm 5mm, clip, width = 0.49\textwidth]{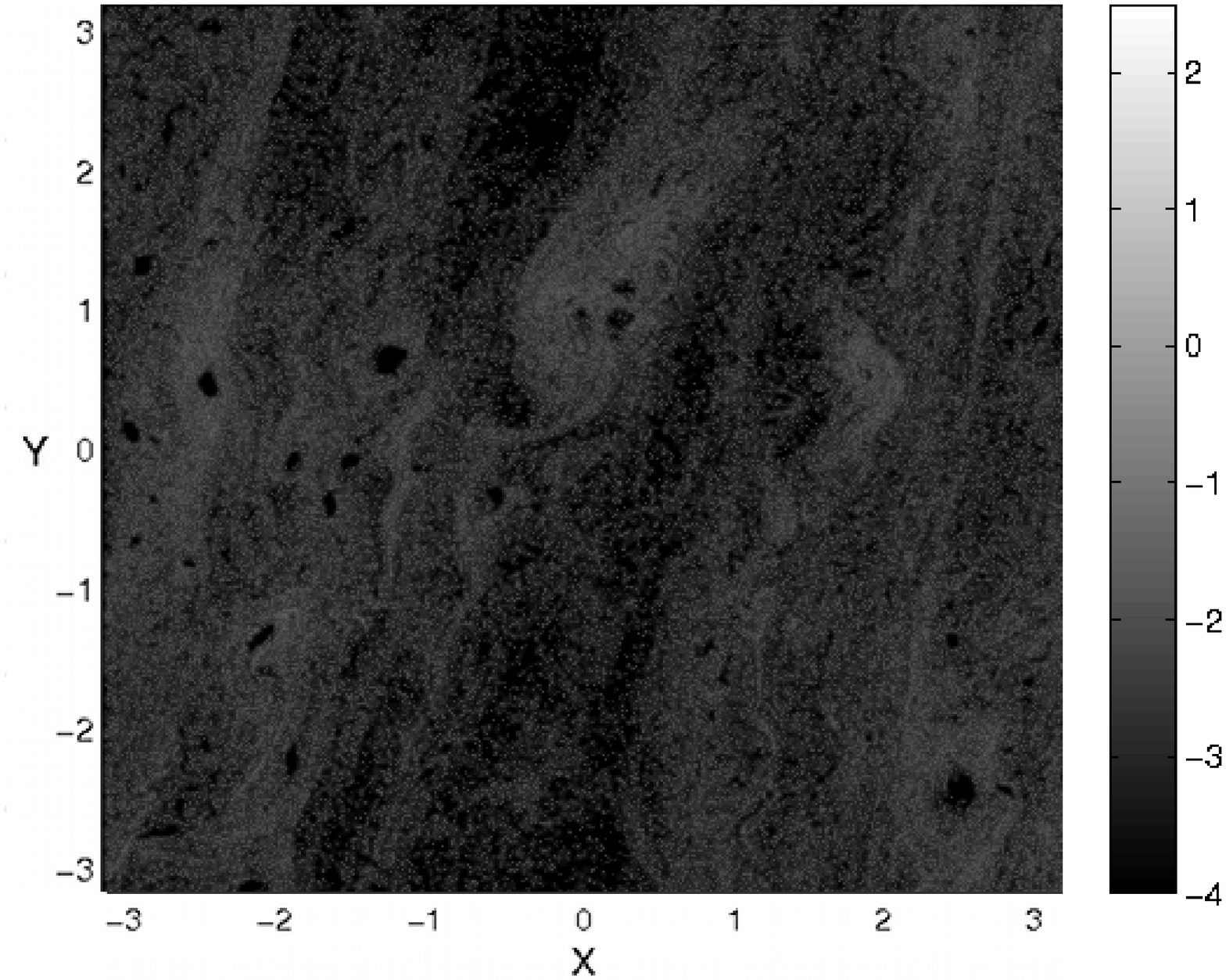}} 
\subfloat{\includegraphics[trim = 10mm 5mm 10mm 5mm, clip, width = 0.49\textwidth]{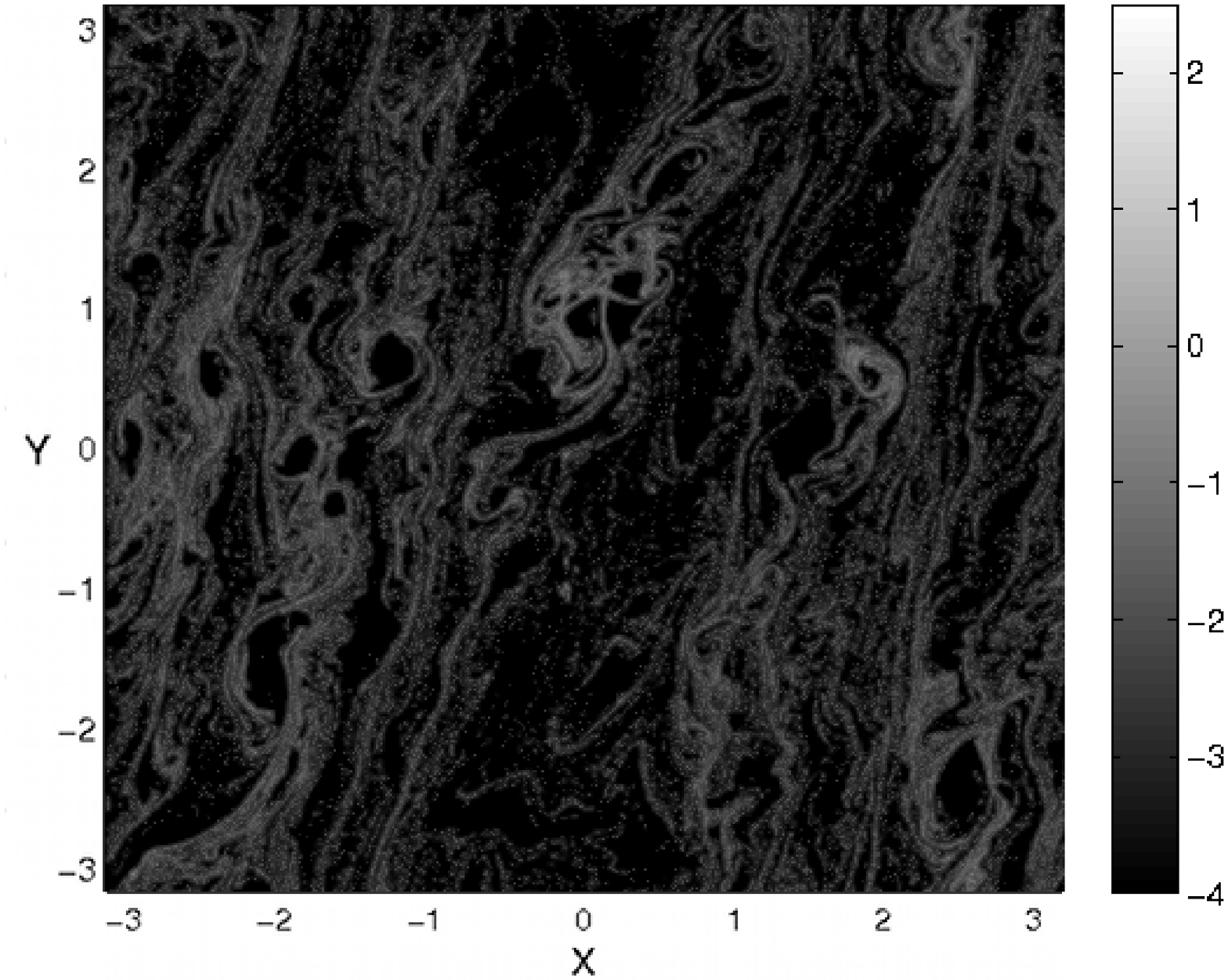}}\\
\subfloat{\includegraphics[trim = 10mm 5mm 10mm 5mm, clip, width = 0.49\textwidth]{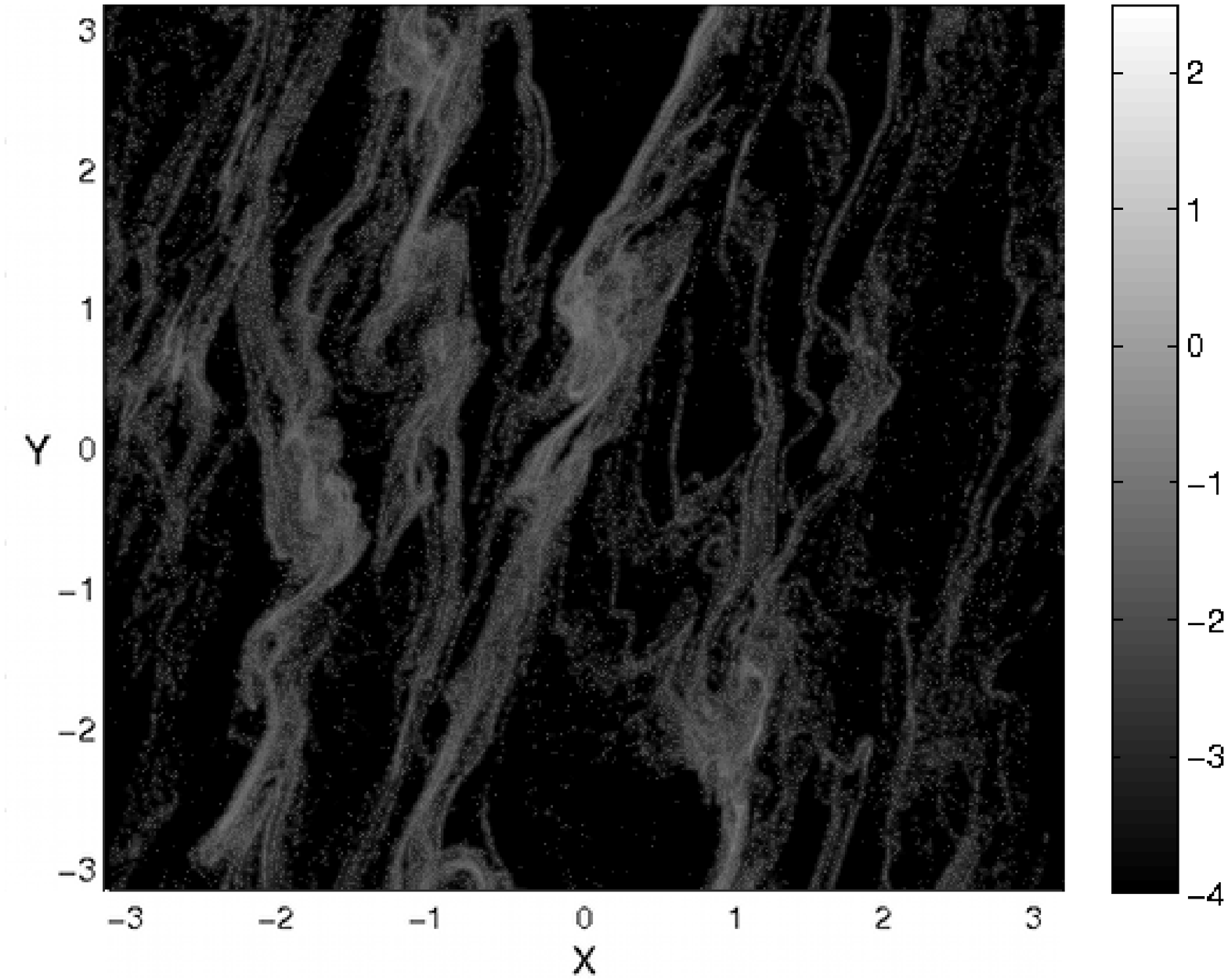}} 
\subfloat{\includegraphics[trim = 10mm 5mm 10mm 5mm, clip, width = 0.49\textwidth]{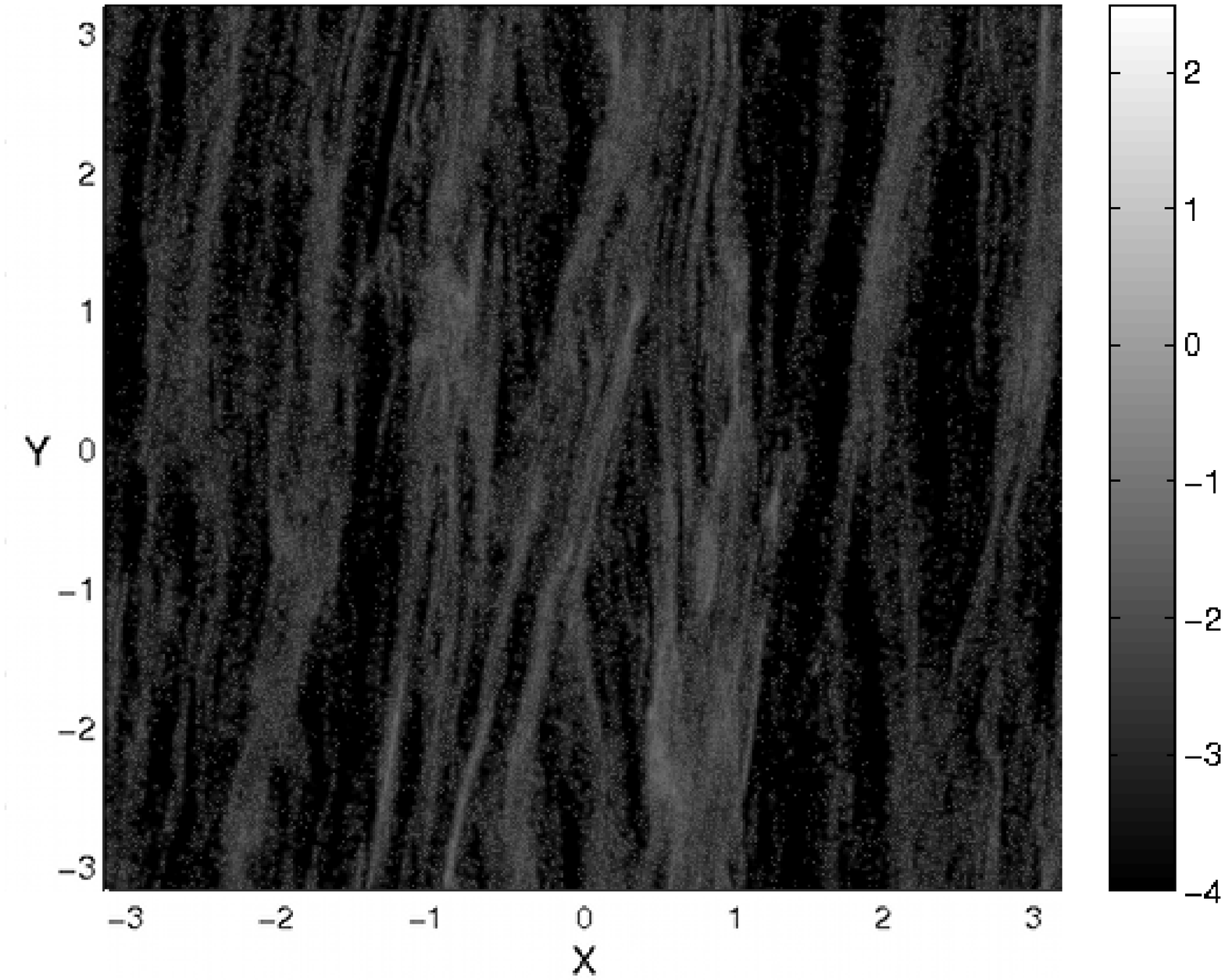}}\\
\subfloat{\includegraphics[trim = 10mm 5mm 10mm 5mm, clip, width = 0.49\textwidth]{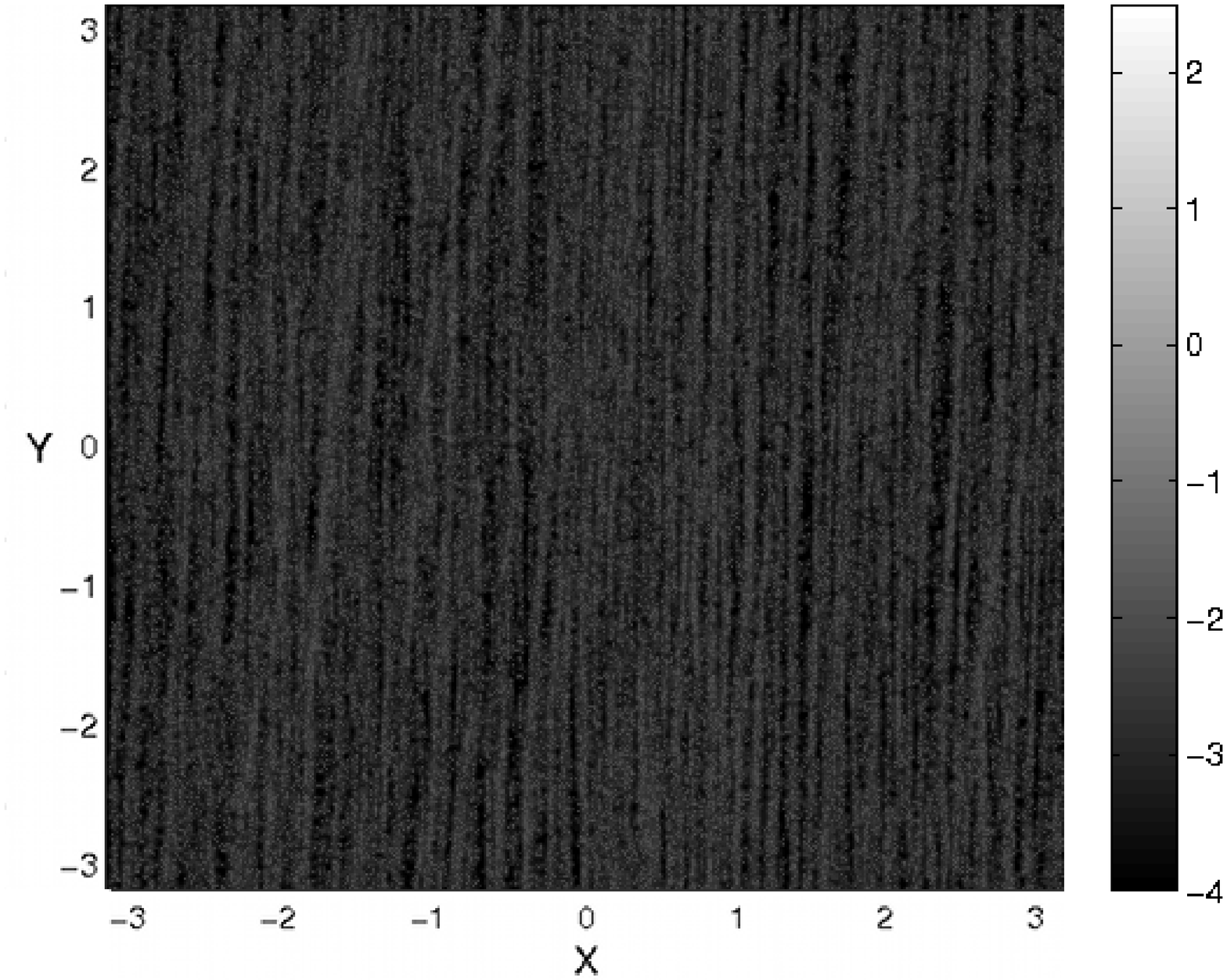}} 
\subfloat{\includegraphics[trim = 10mm 5mm 10mm 5mm, clip, width = 0.49\textwidth]{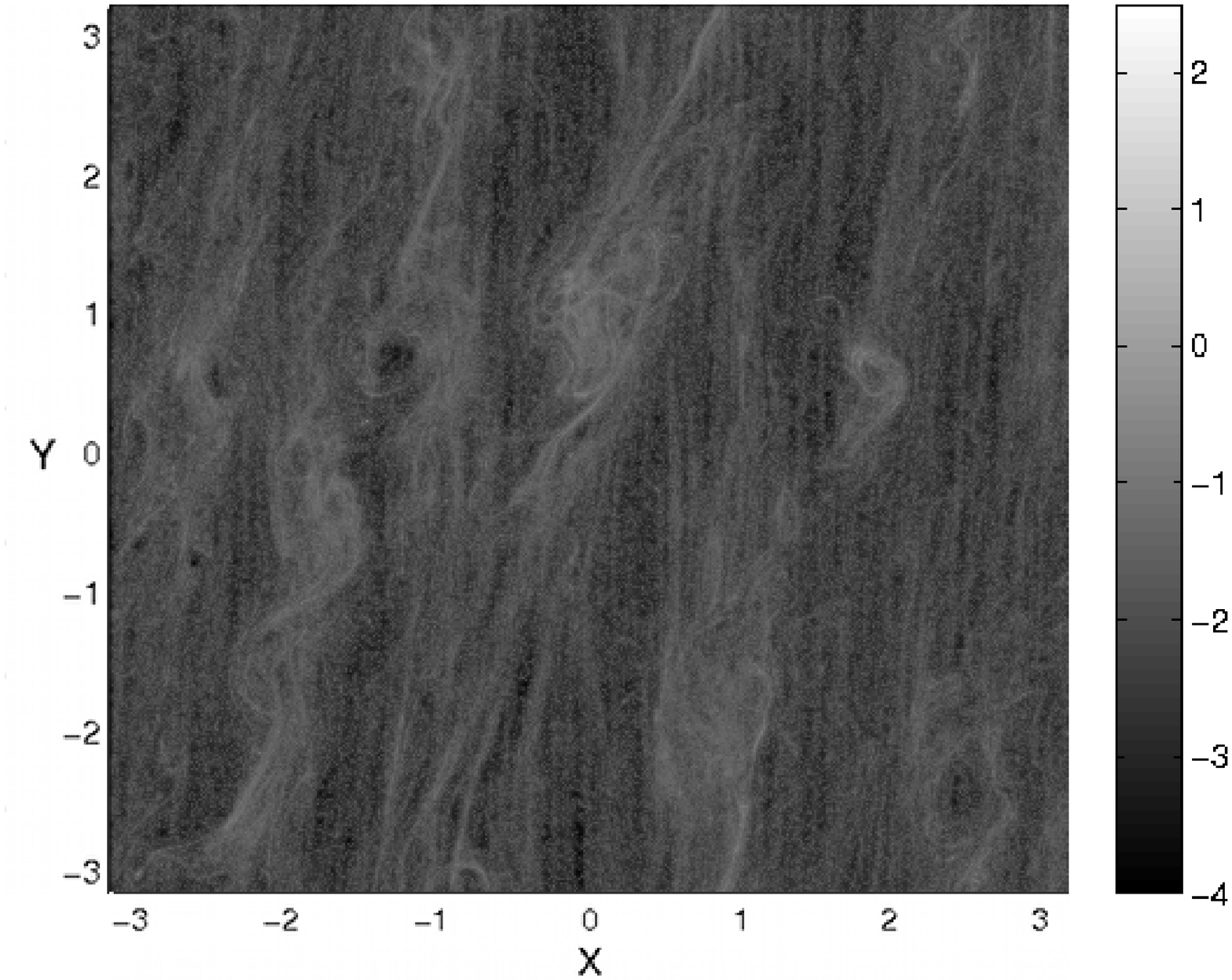}}
\caption{Logarithmic surface density of the individual dust particle species (in units of its mean $\langle \Sigma_p
\rangle=\Sigma_{p0}$) after 30 orbits since the drag force between the gas and particles was introduced. The smallest particles (with
$\tau_f=0.01\Omega^{-1}$) are shown in the top left panel and the other panels, corresponding to increasing friction times, follow in
lexicographic order with the largest particles (with $\tau_f=100\Omega^{-1}$) in the bottom left panel. The total surface
density of the dust is shown in the bottom right panel. For comparison, the surface density of the gas at this time is shown in
Fig. \ref{gas_dens} and the corresponding PV field is shown in Fig. \ref{vor_field}. We observe that large particles with $\tau_f=[10,
100]\Omega^{-1}$ are accumulated in density wave crests; intermediate-sized particles with $\tau_f=1.0\Omega^{-1}$ are
captured effectively in crests of density waves and in the irregularly-shaped anticyclonic vortices, whilst the small particles
with $\tau_f=[0.01, 0.1]\Omega^{-1}$ tend to preferentially accumulate around anticyclonic vortices but do not fill their
central regions, as indicated by black voids, and trace out their structure (see Fig. \ref{vor_field}).}\label{surf_density}
\end{figure*}

In this quasi-steady gravitoturbulent state, the vortices are
transient/unsteady structures undergoing recurring phases of
formation, growth to sizes comparable to a local Jeans scale and
eventual shearing and destruction due to the combined effects of
self-gravity (gravitational instability) and background Keplerian
shear. Each phase typically lasts about two orbital periods or less.
As a result, in self-gravitating discs, the overall dynamical
picture of vortex evolution is irregular consisting of many
transient vortices at different evolutionary stages and, therefore,
with various sizes up to the local Jeans scale. By contrast, in the
non-self-gravitating case, long-lived vortex structures persist for
hundreds of orbits via merging of smaller vortices into larger ones
until eventually their size reaches the disc scale height.

It should be noted that the disc cooling, which we have described
using a simple constant cooling time prescription, in reality is due
to radiative losses from the disc and depends on its physical
properties \citep[e.g.,][]{Johnson2003,Rafikov2005}. Obviously,
cooling affects disc thermodynamics, which, in turn, through the
baroclinic source, determines vortex (potential vorticity) dynamics.
However, in the presence of self-gravity the situation is somewhat
different. In the case of a constant cooling time, despite being a
simple approximation, the disc settles into the quasi-steady
gravitoturbulent state characterized by a local balance between
heating and cooling, which is qualitatively similar to that with a
more realistic radiative cooling \citep[e.g.,][]{Johnson2003,
Boley2006, Forgan2011}. So, it is expected that the evolution of PV
in the presence of realistic cooling will also be similar to that
with a constant cooling time as long as the disc resides in the
quasi-steady state and the global baroclinic or Rossby wave
instabilities do not intervene in the dynamics.

In non-self-gravitating discs, regular and long-lived anticyclonic
vortices have been shown to be efficient at trapping dust particles
in their cores
\citep[e.g.,][]{Johansen2004,Meheut2012b,Fu2014,Zhu2014}. In Section
\ref{Particle_conc} we examine the particle trapping capabilities of
these unsteady and irregular anticyclonic vortices.

\subsection{Particle Concentration}

\label{Particle_conc} Once the gas has reached a quasi-steady
gravitoturbulent state, the aerodynamic drag force and the
corresponding back-reaction terms are introduced into the particle
and gas evolution, respectively. The simulation is then evolved for
a further 30 orbits, at which time the particles have basically also
settled into a quasi-steady state. The particle self-gravity is then
introduced and the simulation was evolved for another 6 orbits.
After this point, the high particle densities achieved in bound
clumps (see below) result in the self-gravity term becoming dominant
in the particle momentum equation. Consequently, the time step of
the simulation becomes very small such that it is impractical to
evolve the particles further. However, inspection of the particle
density field up to this point indicates that the particles have
been sufficiently evolved to draw some conclusions about their
further behaviour. At this time, particle dynamics is practically no
longer affected by gas and one must also incorporate particle
collisions, which become more significant at high particle
concentrations within each clump, to correctly follow the subsequent
dynamics and contraction of the latter \citep{Johansen2012}.

The surface density of the particles at the end of the run with only
the drag force (i.e., before introducing particle self-gravity) is
plotted in Fig. \ref{surf_density}. As shown in
\citet{Gibbons2012,Gibbons2014}, the small- and intermediate-sized
particles with friction times $\tau_f = [0.01, 0.1, 1.0]\Omega^{-1}$
are efficiently trapped in the density wave structures in the gas,
with the particles of stopping times $\tau_f = 1.0\Omega^{-1}$
exhibiting the highest concentration. Of particular interest for the
present work is the behaviour of the dust particles in the vicinity
of smaller scale vortices that form in and around the density wave
structures. The dynamical behaviour of particles in the vicinity of
these anticyclonic vortices appears to depend on their stopping
time, $\tau_f$. The particles with $\tau_f = [0.01, 0.1]\Omega^{-1}$
tend to accumulate around the central region of anticyclonic
vortices, but do not drift further into their centres, leaving
noticeable voids at the centres, which correspond to the local
minima of PV (Fig. \ref{vor_field}). These particles in fact trace
out the over-dense structures around vortices, which, as a result,
appear to be surrounded by a spiral, or ring, of dust particles.
These rings also tend to contain some of the highest concentrations
of particles in the domain, though not significantly larger than the
concentrations found in the crests of density waves. Thus, the
smaller particles with $\tau_f = [0.01, 0.1]\Omega^{-1}$
preferentially map out the vortical structures. On the other hand,
the intermediate-sized particles with friction times
$\tau_f=1.0\Omega^{-1}$ get trapped into crests of density waves and
even more in negative PV regions with relatively low by absolute
value PV (curly areas in Fig. \ref{vor_field}), which produce higher
over-densities in the gas (Fig. \ref{gas_dens}). As opposed to the
smaller friction time case, these particles fill the central parts
of over-dense regions and do not leave voids. The larger particles
with $\tau_f = [10.0, 100.0]\Omega^{-1}$ accumulate only in density
wave crests; vortices do not seem to affect their motion, as is
clear from the surface density of these particles.

\begin{figure}
  \includegraphics[trim = 0mm 0mm 0mm 0mm, clip, width = 0.5\textwidth]{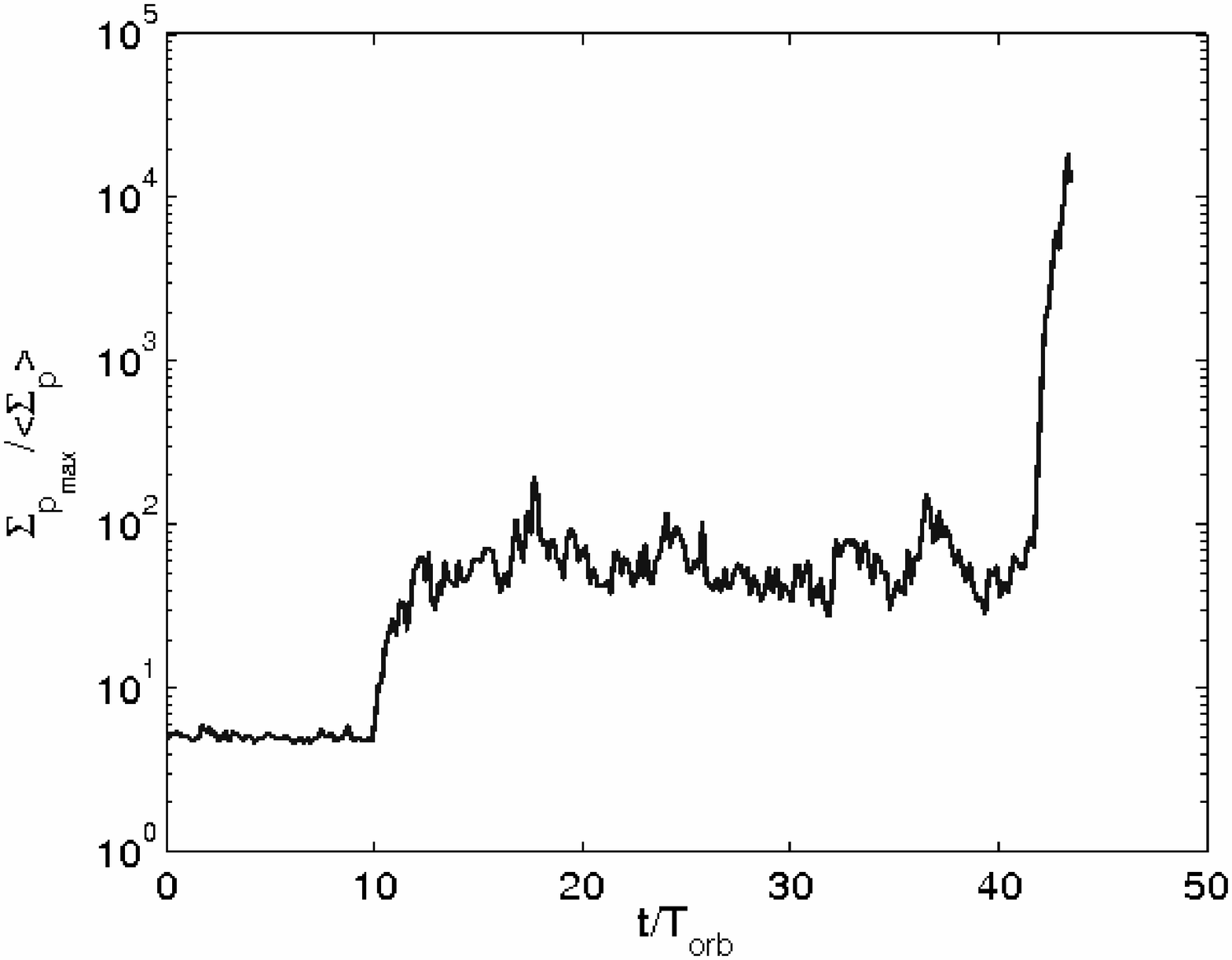}
  \caption{Maximum surface density of the particles within
  the domain as a function of time. The jump in the density at $t=10T_{orb}$
  corresponds to the introduction of the drag force between the gas and dust
  particles, whilst the second jump indicates the introduction of the particle self-gravity.}
  \label{Sigma_t}
\end{figure}

This differential concentration of particles in the vortical
structures depending on their friction time can be explained as
follows. As described above, these small-scale vortices are
transient structures, which undergo cycles of formation,
amplification and shearing away over a few orbits
\citep{Mamatsashvili2009}. The small particles, with friction times
less than the orbital time, are tightly coupled to the gas and
therefore closely follow the density variations induced by vortices.
During short times, these transient vortices are approximately in
balance by self-gravity, pressure and rotation, forming under-dense
core, coinciding with the minimum of PV, surrounded by over-dense
ring. As a result, the small particles, rapidly adjusting to the
vortical motions, get mostly concentrated in these over-dense rings,
leaving the vortex centre practically empty. On the other hand,
during orbital time the vortices evolve, their PV reduces by
absolute value and the corresponding over-density becomes stronger
and no longer has an under-dense core. The particles with friction
times comparable to the orbital time tend to accumulate in these
over-densities repeating their shape, though with a little delay.
Larger particles are not affected by these transient vortices,
because their friction times are longer than the orbital time -- the
characteristic time-scale of vortices.

\begin{figure*}
  \subfloat{\includegraphics[trim = 10mm 5mm 10mm 5mm, clip, width = 0.49\textwidth]{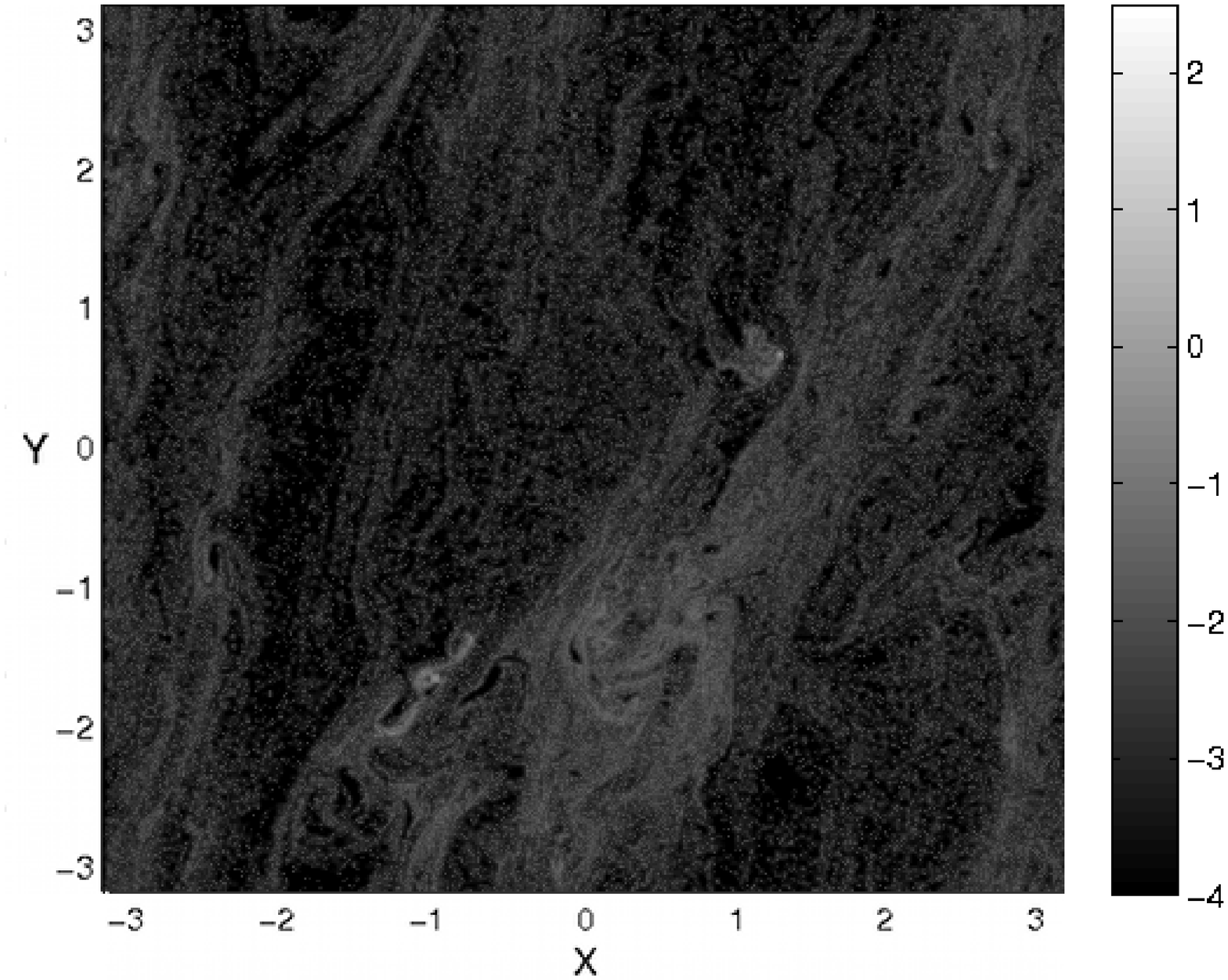}}
  \subfloat{\includegraphics[trim = 10mm 5mm 10mm 5mm, clip, width = 0.49\textwidth]{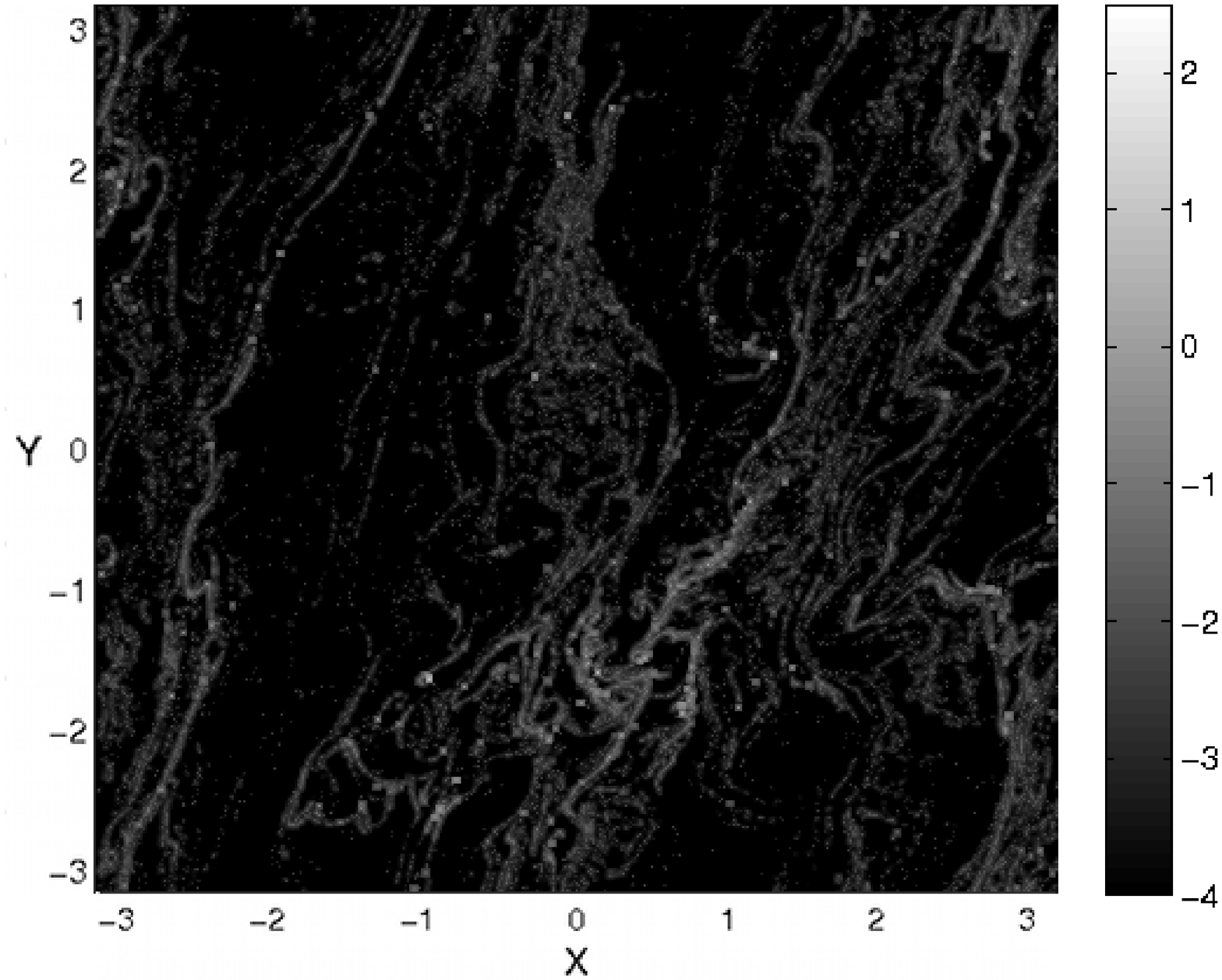}}\\
  \subfloat{\includegraphics[trim = 10mm 5mm 10mm 5mm, clip, width = 0.49\textwidth]{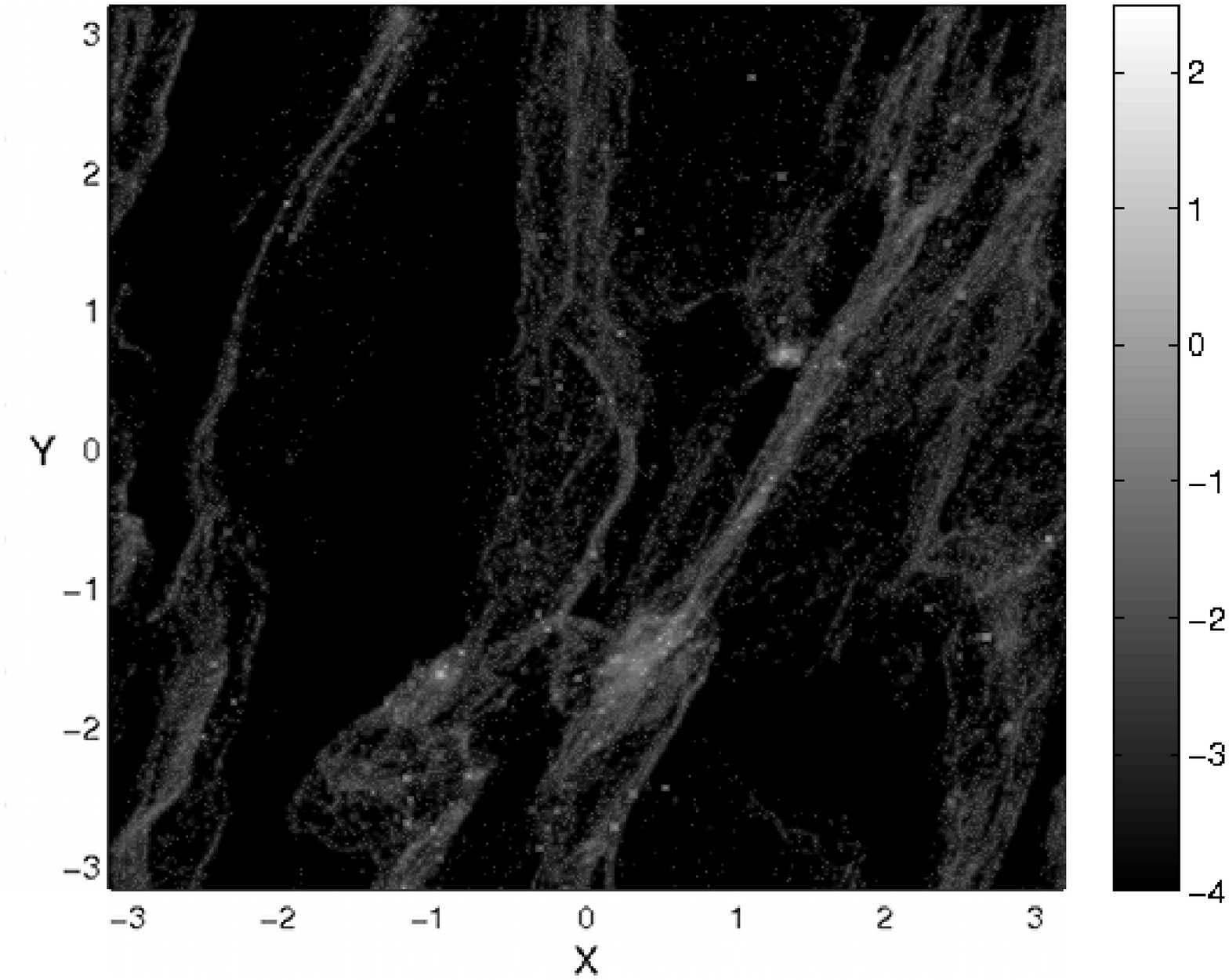}}
  \subfloat{\includegraphics[trim = 10mm 5mm 10mm 5mm, clip, width = 0.49\textwidth]{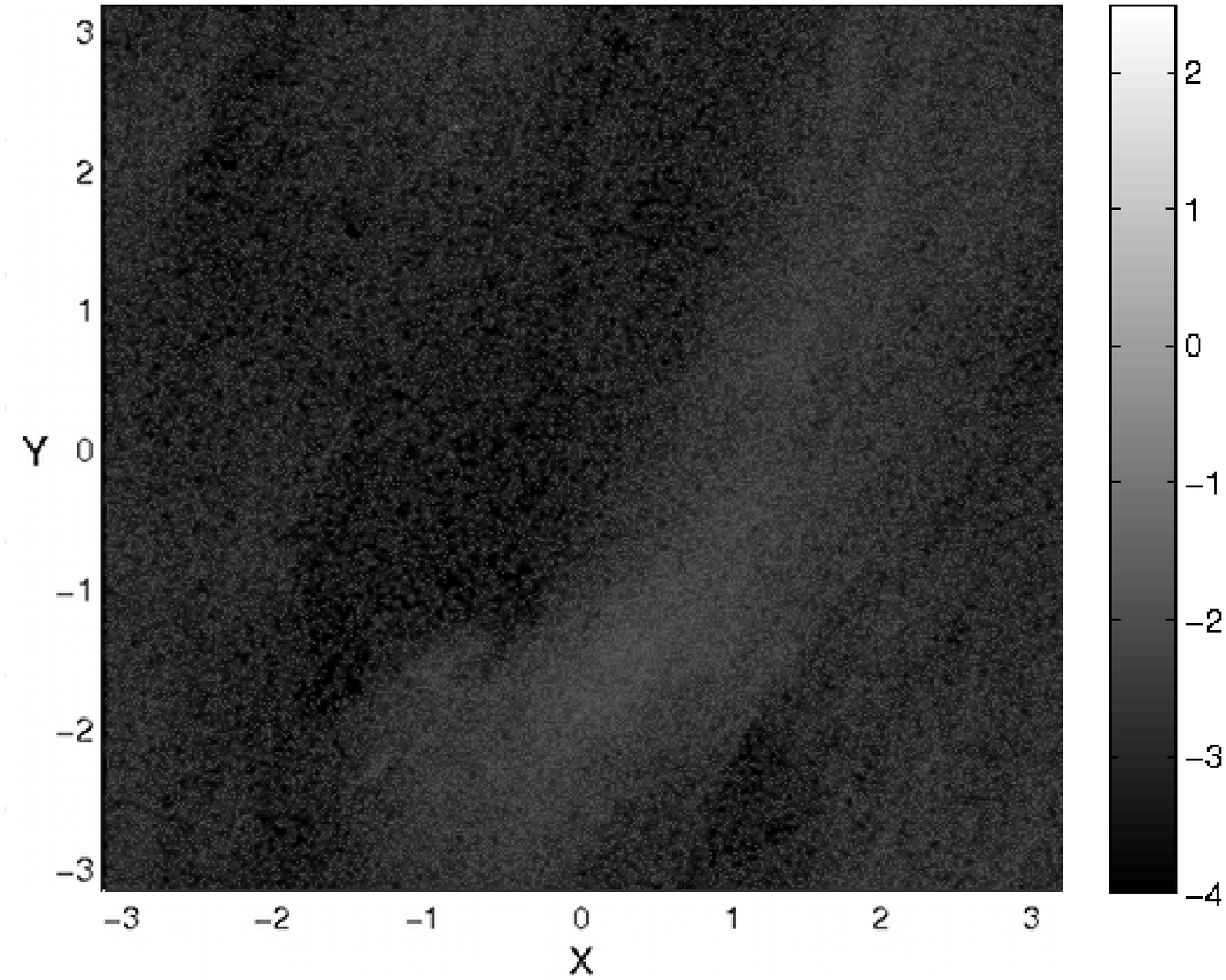}}\\
  \subfloat{\includegraphics[trim = 10mm 5mm 10mm 5mm, clip, width = 0.49\textwidth]{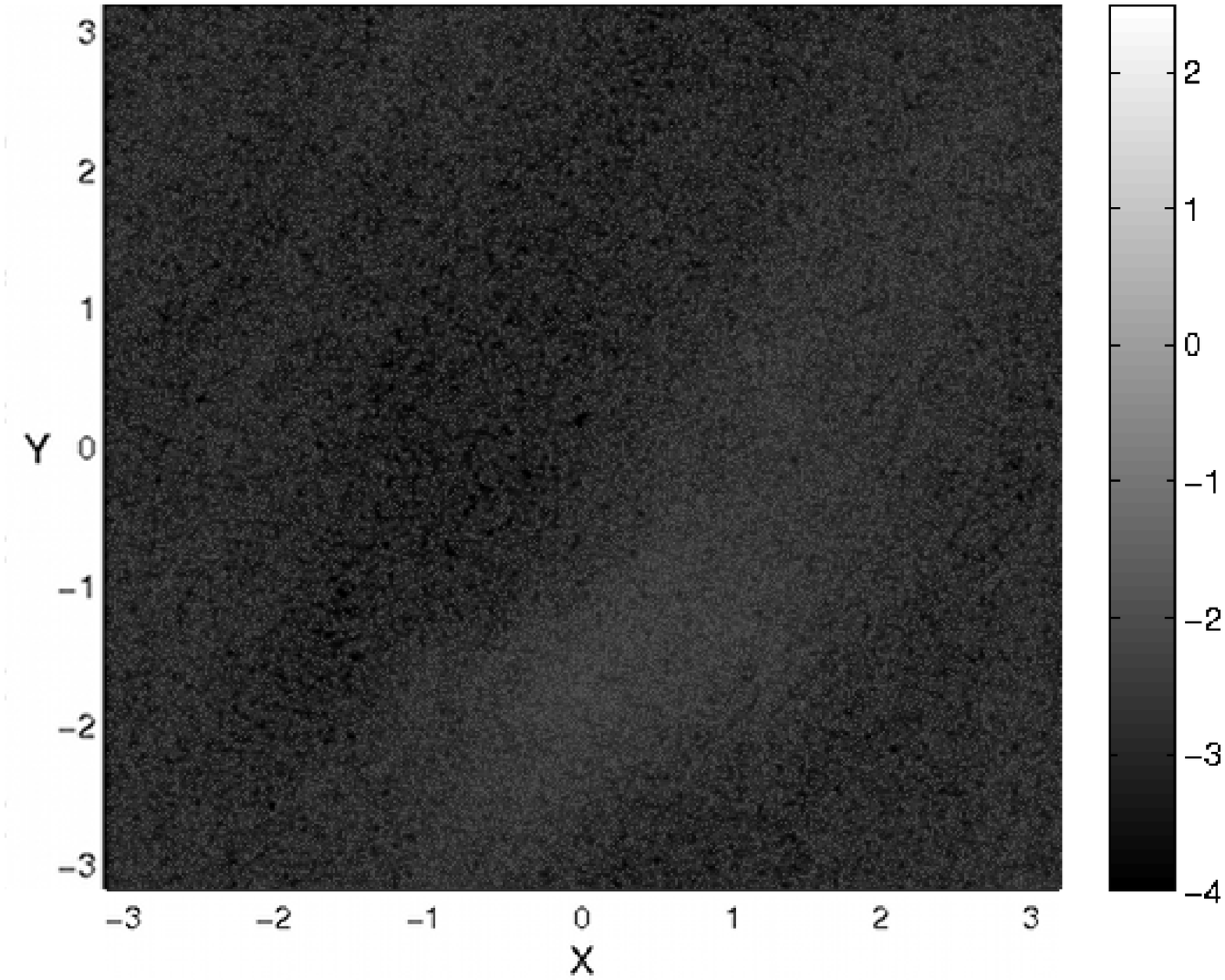}}
  \subfloat{\includegraphics[trim = 10mm 5mm 10mm 5mm, clip, width = 0.49\textwidth]{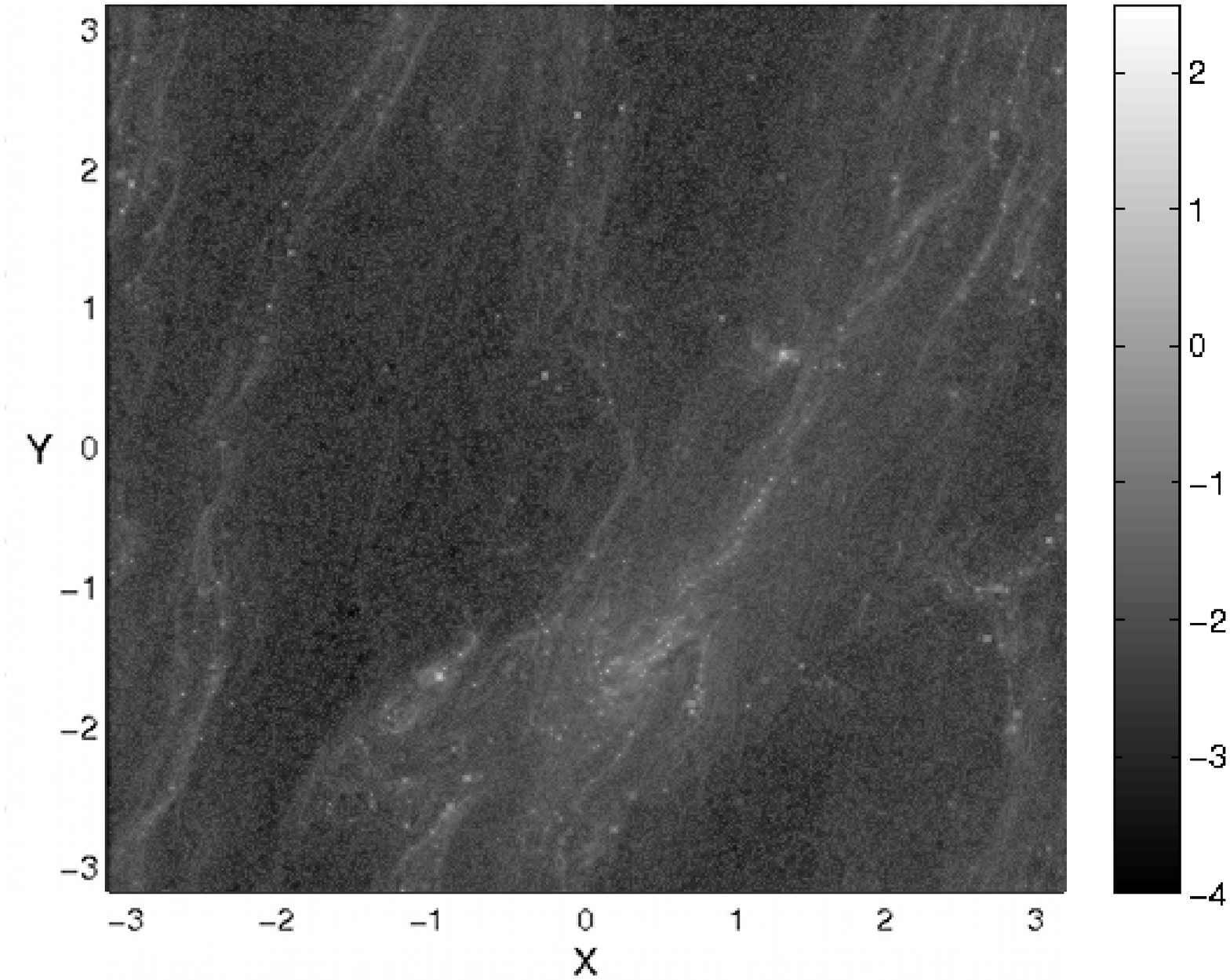}}
  \caption{Logarithmic surface density of the individual particle species (in units of its mean $\langle \Sigma_p \rangle=\Sigma_{p0}$) as well
as the total surface density (bottom right panel) of the dust at the
end of the simulation. Arrangement of the panels according to
stopping times is same as in Fig. \ref{surf_density}. At this time,
the surface density of gas and the PV field are shown, respectively,
in Figs. \ref{gas_dens_sg} and \ref{vor_field_sg}. Only particles
with $\tau_f=[0.1, 1.0]\Omega^{-1}$ experience rapid gravitational
collapse primarily inside the irregular anticyclonic vortices
(compare with Figs. \ref{gas_dens_sg} and \ref{vor_field_sg}). In
such vortices, particles of this size tend to accumulate most
effectively and reach relatively high concentrations (see also Fig.
\ref{surf_density}), leading to the formation of many
gravitationally bound particle aggregates, or clumps (white dots)
dense enough to survive long after the gas over-density they formed
in has dispersed.}
  \label{surf_density_sg}
\end{figure*}

\begin{figure}
  \includegraphics[trim = 10mm 5mm 10mm 5mm, clip, width = 0.5\textwidth]{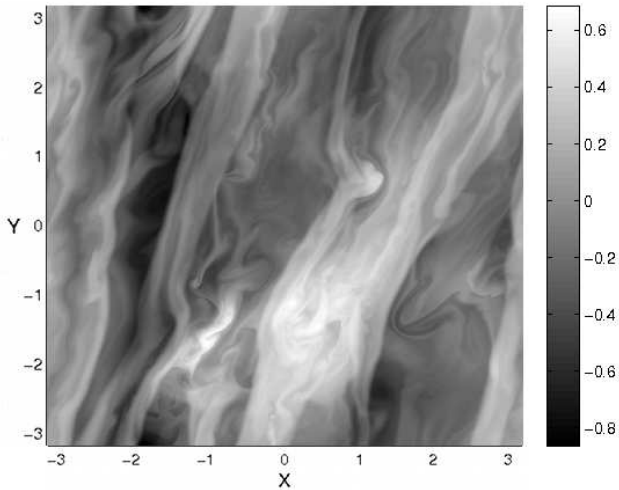}
  \caption{Logarithmic surface density of the gas at the end of the simulation.}
  \label{gas_dens_sg}
\end{figure}

\begin{figure}
  \includegraphics[trim = 10mm 5mm 10mm 5mm, clip, width = 0.5\textwidth]{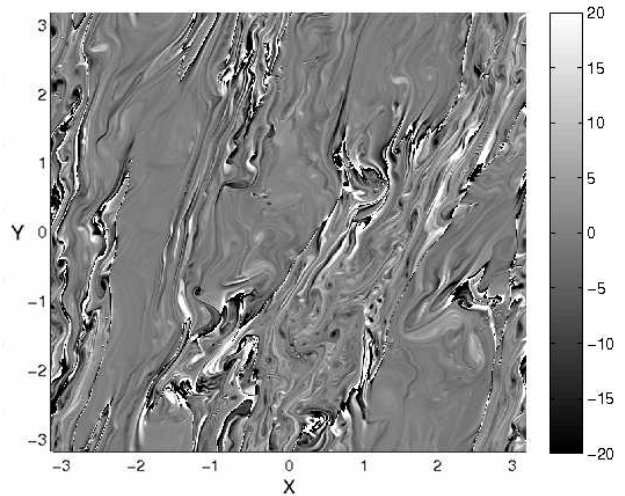}
  \caption{PV field at the same time as in
  Fig. \ref{gas_dens_sg}, which resembles that in Fig. \ref{vor_field}.
  The colour-map is restricted again to the range $-20\leq I\leq 20$.
  The traces of the PV distribution are clearly seen in
  the surface density maps of gas and particles at this time depicted in
  Figs. \ref{gas_dens_sg} and \ref{surf_density_sg}, respectively.}
  \label{vor_field_sg}
\end{figure}

Figure \ref{Sigma_t} shows the evolution of the maximum surface
density (relative to the mean) of solids in the simulation domain.
When the drag force between the particles and the gas is introduced
after 10 orbits, the maximum surface density rapidly jumps,
increasing by a factor of $\sim 10$, from a few times to almost 100
times the mean over the course of a few orbits. Then, the particles
remain in a quasi-steady state, where they are trapped in vortices
and/or density waves, but these gaseous structures get continuously
sheared away and reappear again over few orbital times, so the
particles leaving a gaseous perturbation are swept up by next one
and so on. This cycle continues until the self-gravity of the
particles is introduced after 40 orbits. At this point, the particle
concentrations with large enough densities that have formed before
this time start to undergo rapid gravitational collapse, forming
many gravitationally bound clumps of particles, which undergo
further mergers. The largest of these objects exhibit local surface
densities thousands of times the mean surface density. These objects
are sufficiently stable to remain bound after the parent density
wave and/or vortical perturbation they formed in had been sheared
away. The boundedness of the clumps is established by calculating
total, kinetic plus gravitational energies inside an individual
clump and, if the latter is negative, the clump is classified as
gravitationally bound. The map of the surface density of the solids
at the end of the simulation is shown in Fig. \ref{surf_density_sg},
where we clearly see bound clumps of particles formed due to their
own self-gravity as white dots. These clumps are mostly made up of
intermediate-sized particles with $\tau_f=[0.1, 1.0]\Omega^{-1}$,
which tend to concentrate most effectively (see also Fig.
\ref{surf_density}). The masses of these clumps are typically of the
order of $10^{-3}-10^{-2}M_{Earth}$ \citep{Gibbons2014}. At this
time, the gas density and PV fields are shown in Figs.
\ref{gas_dens_sg} and \ref{vor_field_sg}, respectively. Remarkably,
majority of these clumps are formed within the vortices. Consistent
with the above case before switching on particle self-gravity (Fig.
\ref{surf_density}), the particles with $\tau_f=0.1\Omega^{-1}$
experience clumping in over-dense rings around vortices, whereas
larger particles with $\tau_f=1.0\Omega^{-1}$ in more diffuse
vortices with negative and smaller by absolute value PV (curly areas
in Fig. \ref{vor_field_sg}). As mentioned above, these evolved
vortices tend to give rise to higher over-densities in the gas than
the density waves do, as evident from Fig. \ref{gas_dens_sg}
(especially the one in the central-lower region). Thus, irregular,
smaller vortices appear to be more efficient at trapping particles
than relatively large scale density waves. This suggests that
vortical perturbations can be as important as density waves in
accumulating solids in self-gravitating discs.

\section{Summary and discussion}

In this paper, we investigated the particle trapping properties of
vortices and the potential role they can play in the planet
formation process at the early stage of protoplanetary disc life, when most
of its interior is dense and cool and self-gravity is a dominant
agent governing the dynamics and evolution of the disc. However,
mass accretion due to the gravitational instability can raise the
temperature and activate the magnetorotational instability in the
inner disc \citep[e.g.,][]{Zhu2009, Zhu2010, Rice2009}, but this
effect is beyond the scope of the present analysis, which focuses on
the effects of the gravitational instability only. We used a similar
simulation set-up as adopted in our previous work
\citep{Gibbons2012,Gibbons2014}, but took a smaller simulation
domain, which allowed us to capture small-scale vortical structures
against a backdrop of larger scale density waves. The large-scale
picture of the early stages of planetesimal and planet formation
described in the above papers remains unchanged: dust particles
accumulate within a quasi-steady gaseous spiral structure produced
by a combined effect of disc self-gravity and cooling. The work
performed here gives us deeper insight into how smaller scale
processes affect the dynamics and evolution of particles in the
quasi-steady gravitoturbulence. In this state, the smaller scale
turbulent vortical motions (eddies) leave their traces in the
evolution of the gas density by producing over-dense regions along
with spiral density waves. Particles are accumulated both in crests
of density waves and in the anticyclonic irregularly-shaped unsteady
vortices. Vortices with small, by absolute value, PV are
characterized by higher over-densities than those with larger, by
absolute value, PV and consequently the particle concentrations in
the former turn out to be accordingly higher, even larger than in
the density waves for smaller and intermediate-sized particles (Fig.
\ref{surf_density}). When the self-gravity of the solid component of
the disc is introduced, the particle over-densities that form within
these irregular vortices and wave crests rapidly undergo
gravitational collapse, forming bound clumps with local densities
hundreds of times larger than the typical gas densities within the
sheet. Remarkably, the number of clumps formed in the vortices is
somewhat larger than that formed in the wave crests (Fig.
\ref{surf_density_sg}). It is noteworthy that the degree of particle
concentration in these structures and hence their clumping property
depend on their friction time. We found that small and
intermediate-sized dust particles corresponding to friction times
comparable to, or less than, the local orbital period of the disc,
$\tau_f \le 1.0\Omega^{-1}$, become especially tightly trapped. This
suggests that vortical perturbations, which generally are an
essential ingredient, together with density waves, in the dynamics
of self-gravitating discs can also play an important role in the
trapping of dust particles and planetesimal formation at early
stages of disc life.

We would like to stress once again that the principal reason for
high densities and clumping of particles occurring in this problem
is their self-gravity. Here, we have not included particle
collisions and one might ask how this effect could modify the
results. \cite{Johansen2012} investigated the effect of collisions
on the particle dynamics in gaseous discs. They found that
collisions, which become more important at high particle
concentrations, actually tend to promote particle concentration via
damping their rms velocities: the maximum densities of particles
including collisions turned out to be more than a factor of three
higher compared to those without collisions. The collisions also
seem to have no apparent effect on the further clumping/collapse
process of these particle concentrations due to their own
self-gravity. The masses of the most massive particle clumps
(planetesimals) formed are relatively insensitive to the presence or
absence of collisions. Based on these results, we anticipate that
inclusion of collisions in our problem would not much alter a basic
dynamical behavior of particles, although extension of this study
with treatment of collisions between numerical super-particles
\citep[for example, such as in][]{Johansen2012} is needed to support
this conclusion.


In this study, we have considered the simplified case of a
razor-thin (2D) disc with a simple cooling law in order to gain
first insight into the effects of vortices on dust particles in the
presence of self-gravity. Obviously, for a fuller understanding,
three-dimensional (3D) treatment of vortices, with embedded
particles, in self-gravitating discs is necessary, which, as far as
we are aware, has not been done yet. The dynamics of 3D vortices in
non-self-gravitating discs has been investigated in a number of
studies \citep[e.g.,][]{Barranco2005, Shen2006, Lesur2009,
Meheut2012a, Richard2013}. In the 3D case, in contrast to the 2D
one, they are subject to the elliptical instability, which
eventually destroys vortices. However, stratification and
baroclinic/Rossby wave driving can counteract the elliptical
instability, saving vortices from destruction. A self-gravitating 3D
disc is clearly stratified in the vertical direction. On the other
hand, self-gravity does not favour long-lived regular vortices. But
in the 3D case, the effect of self-gravity is somewhat reduced
compared to that in the 2D case \citep[e.g.,][]{Goldreich1965,
Mamatsashvili2010}. Thus, it remains to be seen in future numerical
studies how the combination of the above factors shapes the dynamics
of 3D vortices in self-gravitating discs. When analyzing behaviour
of particles in such 3D vortices in self-gravitating discs, one must
also take into account the basic effects in the particle dynamics
being at work in 3D -- sedimentation due to vertical gravity and
stir up and clumping of the dust layer due to the Kelvin-Helmholtz
and streaming instabilities, respectively
\citep[e.g.,][]{Youdin2002, Johansen2006, Johansen_Youdin2007,
Johansen2009, Bai2010}.

The work presented here expands on the picture of the early stages
of planetesimal formation built up by
\citet{Gibbons2012,Gibbons2014}, which still presents an attractive
method for the rapid creation of a large reservoir of planetesimals,
along with several massive objects of mass $\sim 10^{-2}M_{Earth}$.
These objects are then likely to continue to grow via the
traditional core accretion process \citep[e.g.,][]{Pollack1996}.
Further work, which allows the replacement of large, gravitationally
bound accumulations of particles with `sink' particles, interacting
with the disc and with each other, could expand the scope of
simulations such as those presented above. This would allow for the
subsequent evolution of these planetesimal-sized objects and could
proved further insight into how quickly the core accretion process
can work in self-gravitating discs.

\section*{Acknowledgments}
This work made use of the facilities of HECToR -- the UKs national
high-performance computing service, which is provided by UoE HPCx
Ltd at the University of Edinburgh, Cray Inc and NAG Ltd, and funded
by the Office of Science and Technology through EPSRCs High End
Computing Programme. GRM acknowledges financial support from the
Rustaveli National Science Foundation (Georgia) grant PG/52/07. WKMR
acknowledges support from STFC grant ST/M001229/1. We thank an anonymous
Reviewer for constructive comments that improved
the presentation of our work.

\appendix

\section{Resolution test}

To investigate the effects of resolution on the dynamics of vortices
in the presence of self-gravity, we ran simulations for a domain
with half size, $L_x/2=L_y/2=10/\pi$ (in units of $H$, initial
$Q=1$) at three different resolutions $N_x\times N_y=256\times 256,
512\times 512, 1024\times 1024$, including the same number of
particles and starting with the same initial conditions as in the
paper. This is equivalent to running the fiducial model with
$L_x=L_y=20/\pi$ considered in this work at doubled resolutions
$512\times 512, 1024\times 1024$ and $2048\times 2048$,
respectively, since the corresponding grid cell sizes, $\delta x=
L_x/N_x=0.0124, 0.0062, 0.0031$ would be the same in both cases. So
far as numerical convergence is concerned, grid cell size is a
central length-scale, since measuring various physical lengths of
the system against it allows us to find out whether a simulation is
resolved. Therefore, due to the same grid cell sizes, a resolution
study we do for the half-size model also establishes whether
vortices in the fiducial model are sufficiently resolved.

\begin{figure}
\includegraphics[width = \columnwidth]{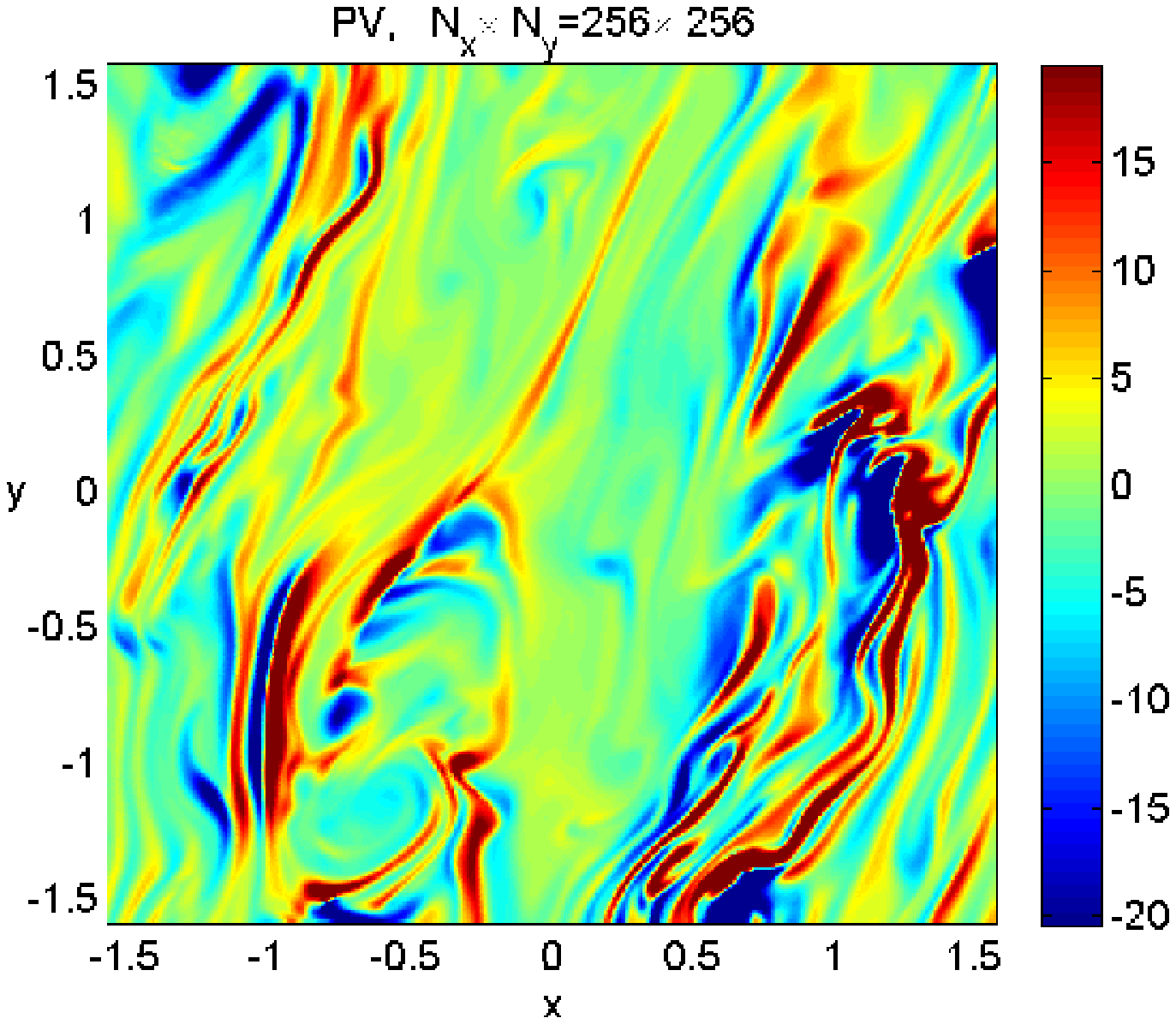}
\includegraphics[width = \columnwidth]{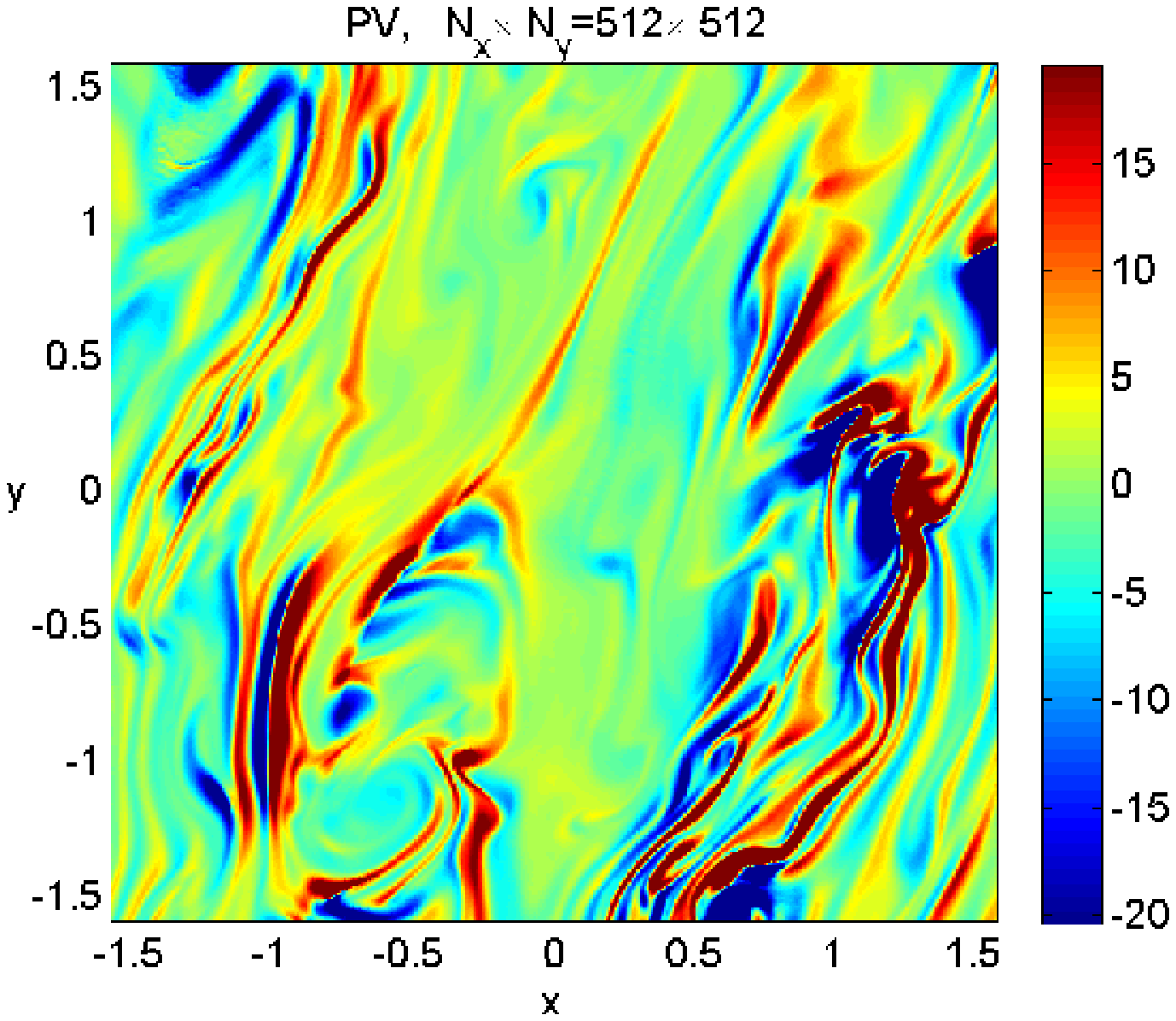}
\includegraphics[width = \columnwidth]{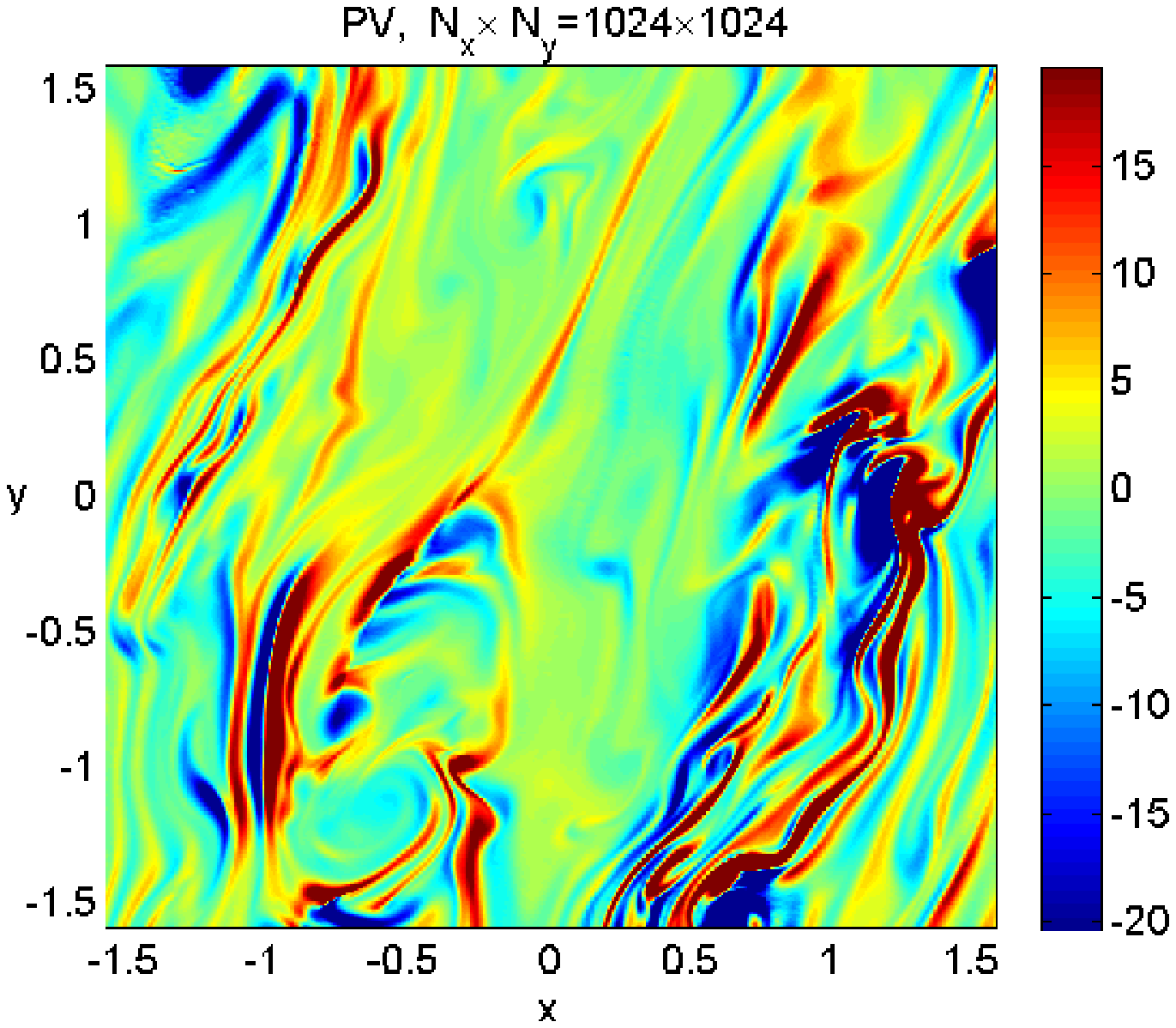}
\caption{PV field for the half size domain at different resolutions in the quasi-steady state at the same time (colour online).}
\end{figure}
\begin{figure}
\includegraphics[width = \columnwidth]{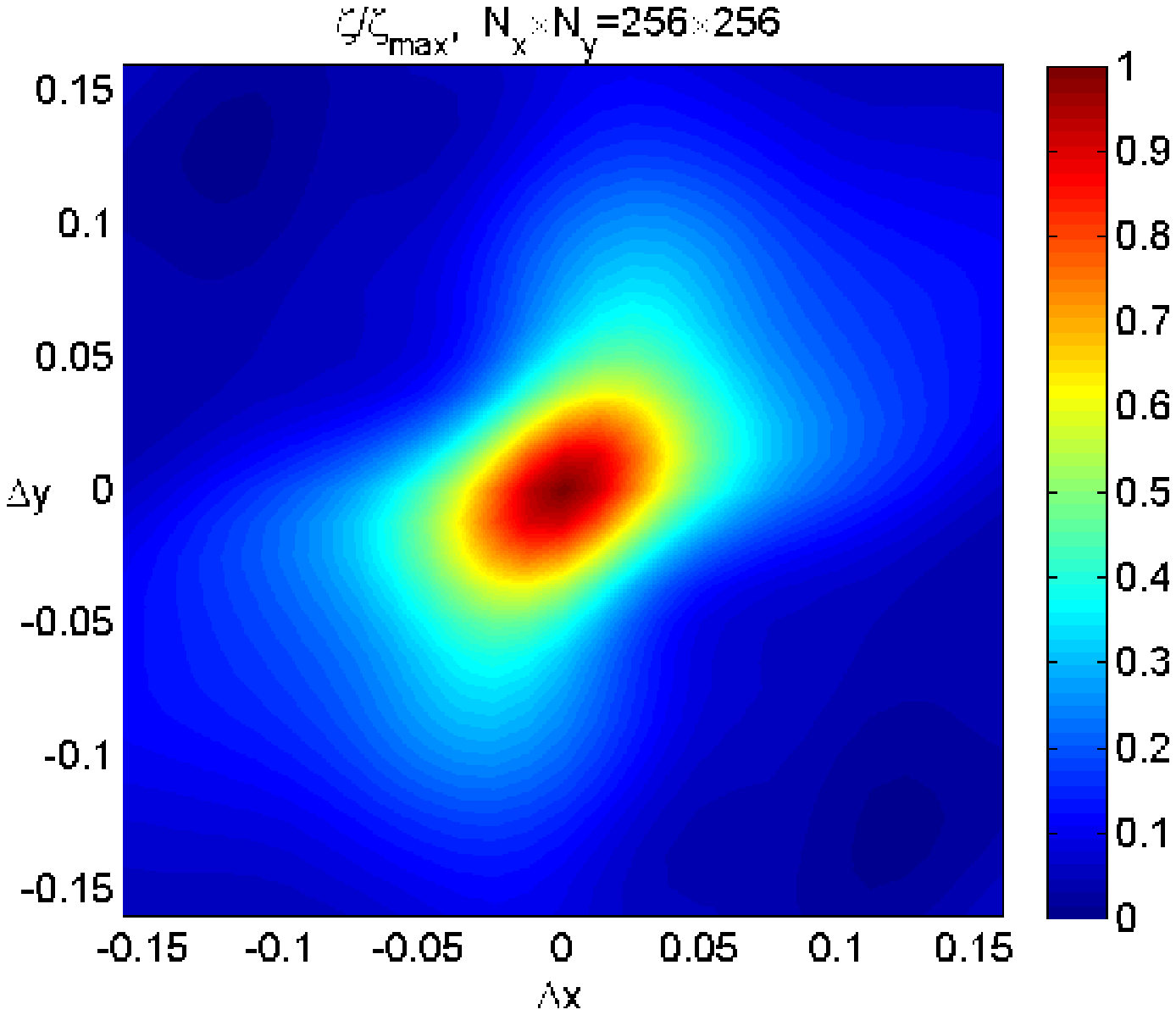}
\includegraphics[width = \columnwidth]{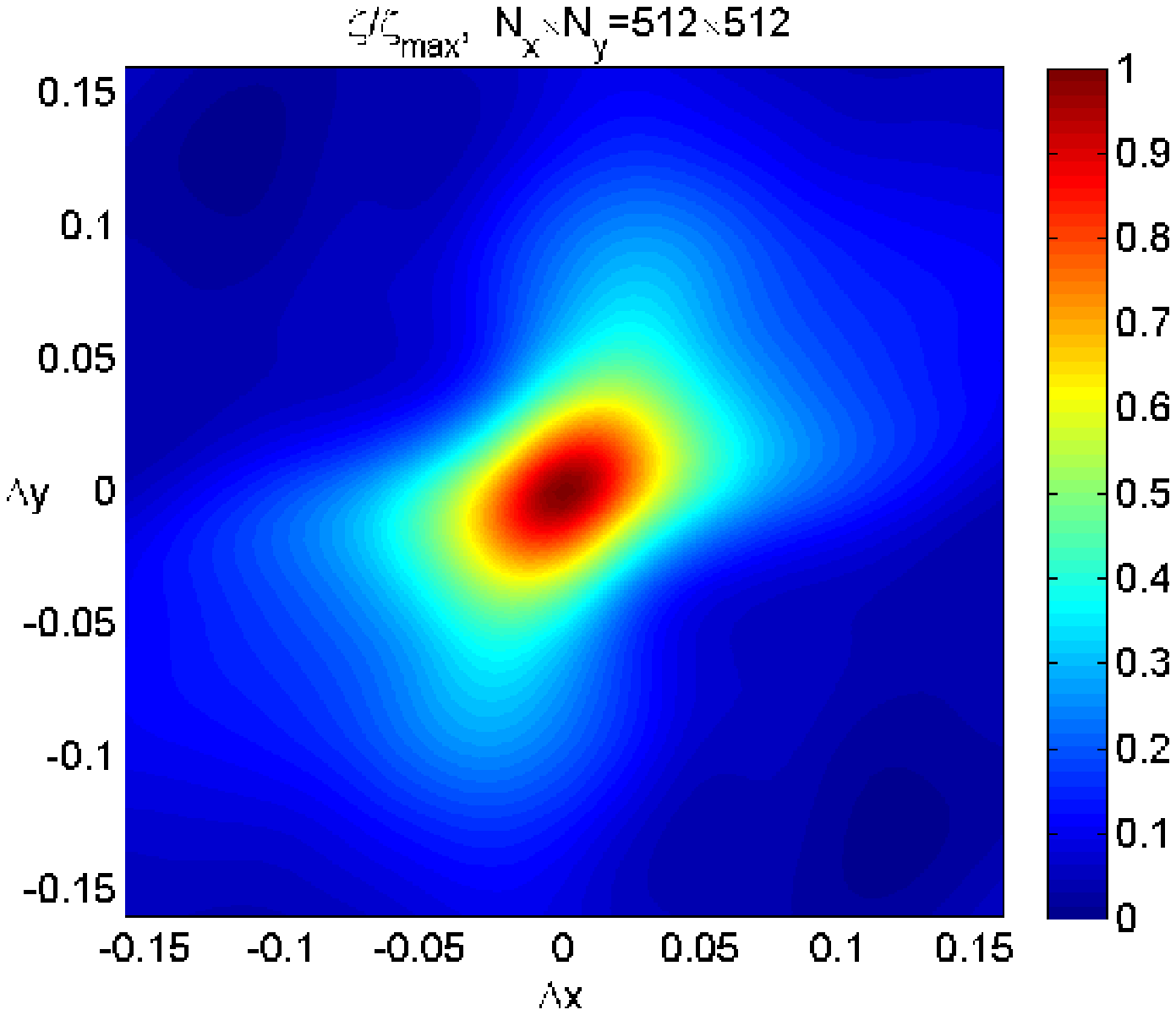}
\includegraphics[width = \columnwidth]{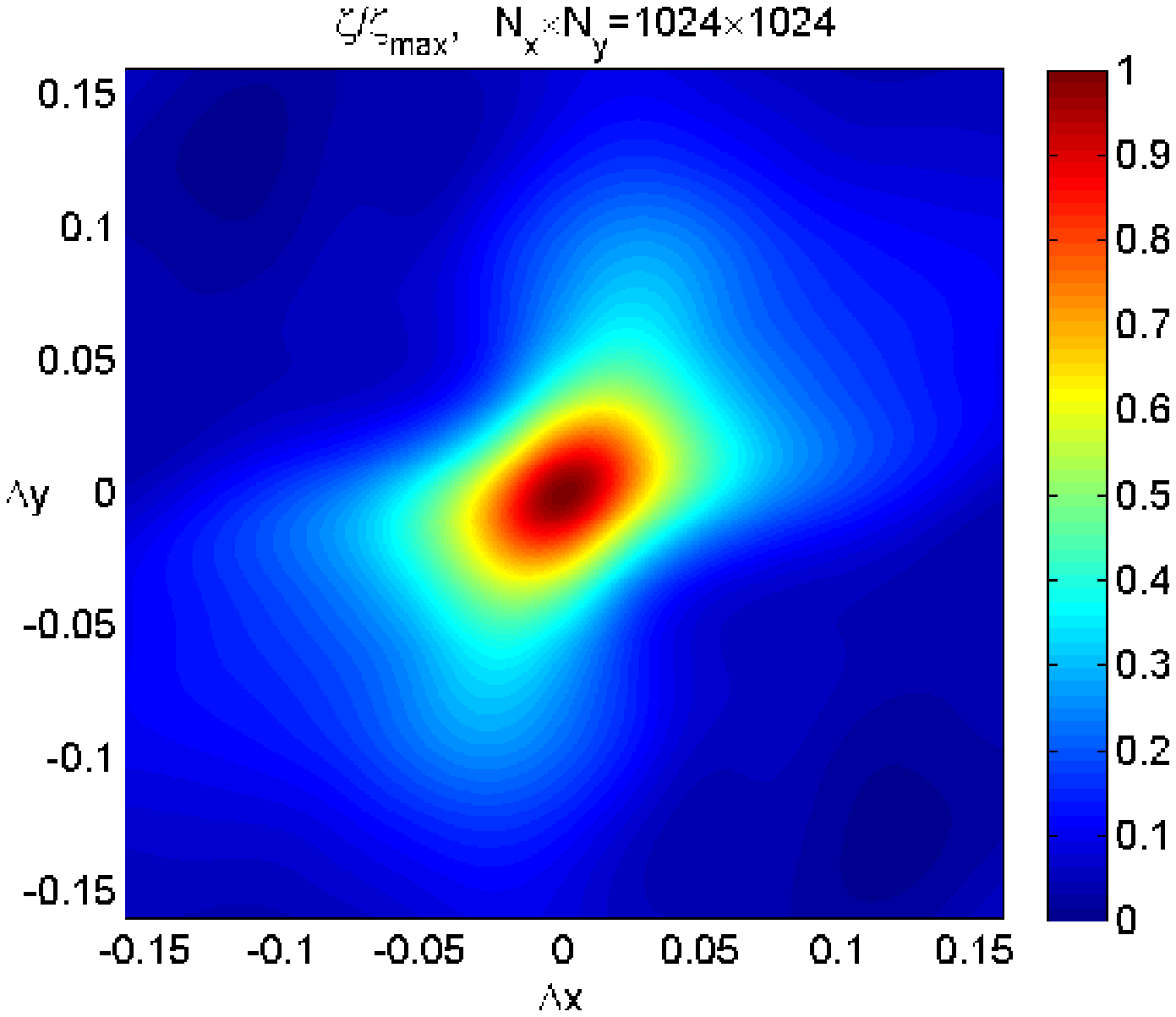}
\caption{Normalized autocorrelation function, $\zeta/\zeta_{max}$, where $\zeta_{max}=\zeta(0)$, corresponding to the PV in Fig. A1 at different resolutions (colour online).}
\end{figure}

In the quasi-steady state, the PV fields from all
these three runs at the same time are shown in Fig. A1.  These fields, being at different resolutions, display essentially
identical structures, indicating that they are resolved. To put this conclusion on more quantitative grounds, as in
\citet{Johnson2005}, we consider the autocorrelation function for the fluctuating part of PV, $\delta I=I-\langle I
\rangle$, given by
\[
\zeta({\bf \Delta x})=\langle \delta I({\bf x})\delta I({\bf x}+{\bf
\Delta x})\rangle,
\]
where the angle brackets denote average over the whole domain.
Figure A2 shows the autocorrelation functions corresponding to the PV fields in Fig. A1. They
do appear similar and their cores have nearly an elliptical shape with
minor and major principal axis. Following the method of \citet{Guan2009}, we
calculated related correlation lengths along these axes by
\[
\lambda_{min}=\frac{1}{\zeta(0)}\int_0^{\infty}\zeta(l\hat{\bf
x}_{min})dl,~~~~\lambda_{maj}=\frac{1}{\zeta(0)}\int_0^{\infty}\zeta(l\hat{\bf
x}_{maj})dl,
\]
where $l$ is the distance from ${\bf \Delta x}=0$ along the
principal axes given by the unit vectors $\hat{\bf x}_{min}$ and
$\hat{\bf x}_{maj}$. These correlation lengths can be used as a
measure for the typical size of the PV structures.

The smallest correlation lengths along the minor axis are
$\lambda_{min}=0.088, 0.079, 0.077$,  for the lowest, intermediate
and highest resolution runs, respectively. The ratios of the minimum
correlation length to the grid cell size at these resolutions are,
respectively, $\lambda_{min}/\delta x=7.1, 12.7, 24.8$. This ratio
increases with resolution which serves as an indication for
convergence and, as a matter of fact, implies that the vortices
(eddies) can be considered to be sufficiently resolved in the
fiducial model which has $\delta x=0.0062$. In our previous papers
\citet{Gibbons2012, Gibbons2014}, the domain size is
$L_x=L_y=80/\pi$ and resolution $1024\times 1024$ for the same other
parameters, giving $\delta x = 0.0249$ and hence
$\lambda_{min}/\delta x<3.5$. Thus, although this resolution is
quite suitable to resolve larger scale density waves in those
studies, it is still coarse to detect and examine vortical
structures at such larger domain sizes in contrast to that at four
times smaller domain size used in the present paper.


\begin{thebibliography}{99}
\bibitem[\protect\citeauthoryear{Andrews \& Williams}{2007}]{Andrews2007} Andrews S. M., Williams J. P., 2007, ApJ, 671, 1800
\bibitem[\protect\citeauthoryear{Ataiee et al.}{2014}]{Ataiee2014} Ataiee S., Dullemond C. P., Kley W., Regaly Z., Meheut H., 2014, A\&A, 572, A61
\bibitem[\protect\citeauthoryear{Bai \& Stone}{2010}]{Bai2010} Bai X.-N., Stone J., 2010, ApJ, 722, 1437
\bibitem[\protect\citeauthoryear{Barge \& Sommeria}{1995}]{Barge1995} Barge P., Sommeria J., 1995, A\&A, 295, L1
\bibitem[\protect\citeauthoryear{Barranco \& Marcus}{2005}]{Barranco2005} Barranco J., Marcus  P., 2005, ApJ, 623, 1157
\bibitem[\protect\citeauthoryear{Bodo et al.}{2007}]{Bodo2007} Bodo G., Tevzadze A., Chagelishvili G., Mignone A., Rossi P., Ferrari A., 2007, A\&A, 475, 51
\bibitem[\protect\citeauthoryear{Bodo et al.}{2005}]{Bodo2005} Bodo G., Chagelishvili G., Murante G., Tevzadze A., Rossi P., Ferrari A., 2005, A\&A, 437, 9
\bibitem[\protect\citeauthoryear{Boley et al.}{2006}]{Boley2006} Boley A. C., Mejia A., Durisen R., Cai K., Pickett M., D'Alessio P., 2006, ApJ, 651, 517
\bibitem[\protect\citeauthoryear{Boss}{1998}]{Boss1998} Boss A. P., 1998, Nature, 393, 141
\bibitem[\protect\citeauthoryear{Brandenburg}{2003}]{Brandenburg2003} Brandenburg A., 2003, in Advances in Nonlinear Dynamos, ed. A. Ferriz-Mas \& M. Nunez (London: Taylor \& Francis), 269
\bibitem[\protect\citeauthoryear{Clarke}{2009}]{Clarke2009} Clarke C.J., 2009, MNRAS, 396, 1066
\bibitem[\protect\citeauthoryear{Cossins et al.}{2009}]{Cossins2009} Cossins P., Lodato G., Clarke C., 2009, MNRAS, 396, 1157
\bibitem[\protect\citeauthoryear{Durisen et al.}{2007}]{Durisen2007} Durisen R. H., Boss A. P., Mayer L., Nelson A. F., Quinn T., Rice, W. K. M., Protostars and Planets V, B. Reipurth, D. Jewitt, and K. Keil (eds.), University of Arizona Press, Tucson, 951, pp., 2007., p.607-622
\bibitem[\protect\citeauthoryear{Eisner et al.}{2005}]{Eisner2005} Eisner J., Hillenbrand L., Carpenter J., Wolf S., 2005, ApJ, 635, 396
\bibitem[\protect\citeauthoryear{Forgan et al.}{2011}]{Forgan2011} Forgan D., Rice W. K. M., Cossins P., Lodato G., 2011, MNRAS, 410, 994
\bibitem[\protect\citeauthoryear{Fu et al.}{2014}]{Fu2014} Fu W., Li H., Lubow S., Li S., Liang E., 2014, ApJ, 795, 39
\bibitem[\protect\citeauthoryear{Gammie}{2001}]{Gammie2001} Gammie C. F., 2001, ApJ, 553, 174
\bibitem[\protect\citeauthoryear{Gibbons et al.}{2014}]{Gibbons2014} Gibbons P. G., Mamatsashvili G. R., Rice W. K. M., 2014, MNRAS, 442, 361
\bibitem[\protect\citeauthoryear{Gibbons et al.}{2012}]{Gibbons2012} Gibbons P. G., Rice W. K. M., Mamatsashvili G. R., 2012, MNRAS, 426, 1444
\bibitem[\protect\citeauthoryear{Godon \& Livio}{2000}]{Godon2000} Godon P., Livio M., 2000, ApJ, 537, 396
\bibitem[\protect\citeauthoryear{Godon \& Livio}{1999}]{Godon1999} Godon P., Livio M., 1999, ApJ, 523, 350
\bibitem[\protect\citeauthoryear{Goldreich \& Lynden-Bell}{1965}]{Goldreich1965} Goldreich P., Lynden-Bell D., 1965, MNRAS, 130, 97
\bibitem[\protect\citeauthoryear{Guan et al.}{2009}]{Guan2009} Guan X., Gammie C., Simon J., Johnson B., 2009, ApJ, 694, 1010
\bibitem[\protect\citeauthoryear{Heinemann \& Papaloizou}{2009}]{Heinemann2009} Heinemann T., Papaloizou J. C. B., 2009, MNRAS, 397, 52
\bibitem[\protect\citeauthoryear{Johansen et al.}{2012}]{Johansen2012} Johansen A., Youdin A., Lithwick Y., 2012, A\&A, 537, A125
\bibitem[\protect\citeauthoryear{Johansen et al.}{2011}]{Johansen2011} Johansen A., Klahr H., Henning Th., 2011, A\&A, 529, A62
\bibitem[\protect\citeauthoryear{Johansen et al.}{2009}]{Johansen2009} Johansen A., Youdin A., Mac Low M.-M., 2009, A\&A, 704, L75
\bibitem[\protect\citeauthoryear{Johansen et al.}{2007}]{Johansen2007} Johansen A., Oishi J. S., Mac Low M., Klahr H., Henning Th., Youdin A., 2007, Nature, 448, 1022
\bibitem[\protect\citeauthoryear{Johansen \& Youdin}{2007}]{Johansen_Youdin2007} Johansen A., Youdin A., 2007, ApJ, 662, 627
\bibitem[\protect\citeauthoryear{Johansen et al.}{2006}]{Johansen2006} Johansen A., Klahr H., Henning T., 2006, ApJ, 643, 1219
\bibitem[\protect\citeauthoryear{Johansen et al.}{2004}]{Johansen2004} Johansen A., Andersen A. C., Brandenburg A., 2004, A\&A, 417, 361
\bibitem[\protect\citeauthoryear{Johnson \& Gammie}{2005}]{Johnson2005} Johnson B.M., Gammie C. F., 2005, ApJ, 635, 149
\bibitem[\protect\citeauthoryear{Johnson \& Gammie}{2003}]{Johnson2003} Johnson B. M., Gammie C. F., 2003, ApJ, 597, 131
\bibitem[\protect\citeauthoryear{Klahr \& Bodenheimer}{2006}]{Klahr2006} Klahr H. H., Bodenheimer P., 2006, ApJ, 639, 432
\bibitem[\protect\citeauthoryear{Klahr \& Bodenheimer}{2003}]{Klahr2003} Klahr H., Bodenheimer P., 2003, ApJ, 582, 869
\bibitem[\protect\citeauthoryear{Lesur \& Papaloizou}{2010}]{Lesur2010} Lesur G., Papaloizou J. C. B., 2010, MNRAS, 513, 60
\bibitem[\protect\citeauthoryear{Lesur \& Papaloizou}{2009}]{Lesur2009} Lesur G., Papaloizou J. C. B., 2009, A\&A, 498, 1
\bibitem[\protect\citeauthoryear{Li et al.}{2001}]{Li2001} Li H., Colgate S. A., Wendroff B., Liska R., 2001, ApJ, 551, 874
\bibitem[\protect\citeauthoryear{Li et al.}{2000}]{Li2000} Li H., Finn J. M., Lovelace R. V. E., Colgate S. A., 2000, ApJ, 533, 1023
\bibitem[\protect\citeauthoryear{Lin \& Pringle}{1990}]{Lin1990} Lin D. N. C., Pringle J. E., 1990, ApJ, 358, 515
\bibitem[\protect\citeauthoryear{Lin \& Pringle}{1987}]{Lin1987} Lin D. N. C., Pringle J. E., 1987, MNRAS, 225, 607
\bibitem[\protect\citeauthoryear{Lin}{2012a}]{Lin2012a} Lin M.-K., 2012a, ApJ, 754, 21
\bibitem[\protect\citeauthoryear{Lin}{2012b}]{Lin2012b} Lin M.-K., 2012b, MNRAS, 426, 3211
\bibitem[\protect\citeauthoryear{Lithwick}{2007}]{Lithwick2007} Lithwick Y., 2007, ApJ, 670, 789
\bibitem[\protect\citeauthoryear{Lovelace \& Hohlfeld}{2013}]{Lovelace2013} Lovelace R. V. E., Hohlfeld R. G., 2013, MNRAS, 429, 529
\bibitem[\protect\citeauthoryear{Lovelace et al.}{1999}]{Lovelace1999} Lovelace R. V. E., Li H., Colgate S. A., Nelson A. F.,  1999, ApJ, 513, 805
\bibitem[\protect\citeauthoryear{Lyra \& Lin}{2013}]{Lyra2013} Lyra W., Lin M.-K., 2013, ApJ, 775, 17
\bibitem[\protect\citeauthoryear{Lyra \& Klahr}{2011}]{Lyra2011} Lyra W., Klahr H., 2011, A\&A, 527, 138
\bibitem[\protect\citeauthoryear{Lyra et al.}{2009}]{Lyra2009} Lyra W., Johansen A., Zsom A., Klahr H., Piskunov N. 2009, A\&A, 497, 869
\bibitem[\protect\citeauthoryear{Lyra et al.}{2008}]{Lyra2008} Lyra W., Johansen A., Khlar H., Piskunov N., 2008, A\&A, 479, 883
\bibitem[\protect\citeauthoryear{Mamatsashvili \& Rice}{2010}]{Mamatsashvili2010} Mamatsashvili G. R., Rice W. K. M., 2010, MNRAS, 406, 2050 \bibitem[\protect\citeauthoryear{Mamatsashvili \& Rice}{2009}]{Mamatsashvili2009} Mamatsashvili G. R., Rice W. K. M., 2009, MNRAS, 394, 2153
\bibitem[\protect\citeauthoryear{Mamatsashvili \& Chagelishvili}{2007}]{Mamatsashvili2007} Mamatsashvili G. R., Chagelishvili G. D., 2007, MNRAS, 381, 809
\bibitem[\protect\citeauthoryear{Meheut et al.}{2012a}]{Meheut2012a} Meheut H., Keppens R., Casse F., Benz W., 2012a, A\&A, 542, 9
\bibitem[\protect\citeauthoryear{Meheut et al.}{2012b}]{Meheut2012b} Meheut H., Meliani Z., Varniere P., Benz W., 2012b, A\&A, 545, 134
\bibitem[\protect\citeauthoryear{Meheut et al.}{2010}]{Meheut2010} Meheut H., Casse F., Varniere P., Tagger M., 2010, A\&A, 516, 31
\bibitem[\protect\citeauthoryear{Paardekooper}{2012}]{Paardekooper2012} Paardekooper S.-J., 2012, MNRAS, 421, 3286
\bibitem[\protect\citeauthoryear{Paardekooper et al.}{2010}]{Paardekooper2010} Paardekooper S.-J., Lesur G., Papaloizou J. C. B., 2010, ApJ, 725, 146
\bibitem[\protect\citeauthoryear{Petersen et al.}{2007}]{Petersen2007} Petersen M. R., Julien K., Stewart G. R., 2007, ApJ, 658, 1236
\bibitem[\protect\citeauthoryear{Pollack et al.}{1996}]{Pollack1996} Pollack J. B., Hubickyj O., Bodenheimer P., Lissauer J., Podolak M., Greenzweig Y., 1996, Icarus, 124, 62
\bibitem[\protect\citeauthoryear{Raettig et al.}{2013}]{Raettig2013} Raettig N., Lyra W., Klahr H., 2013, ApJ, 765, 115
\bibitem[\protect\citeauthoryear{Rafikov}{2005}]{Rafikov2005} Rafikov R. R., 2005, ApJ, 621, L69
\bibitem[\protect\citeauthoryear{Regaly et al.}{2012}]{Regaly2012} Regaly Zs., Juhasz A., Sandor Zs., Dullemond C. P., 2012, MNRAS, 419, 1701
\bibitem[\protect\citeauthoryear{Rice et al.}{2011}]{Rice2011} Rice W. K. M., Armitage P. J., Mamatsashvili G. R., Lodato G., Clarke C. J., 2011, MNRAS, 418, 1356
\bibitem[\protect\citeauthoryear{Rice et al.}{2010}]{Rice2010} Rice W. K. M., Mayo J. H., Armitage P. J., 2010, MNRAS, 402, 1740
\bibitem[\protect\citeauthoryear{Rice \& Armitage}{2009}]{Rice2009} Rice W. K. M., Armitage P. J., 2009, MNRAS, 396, 2228
\bibitem[\protect\citeauthoryear{Rice et al.}{2006}]{Rice2006} Rice W. K. M., Lodato G., Pringle J. E., Armitage P. J., Bonnell I. A., 2006, MNRAS, 372, L9
\bibitem[\protect\citeauthoryear{Rice et al.}{2004}]{Rice2004} Rice W. K. M., Lodato G., Pringle J. E., Armitage P. J., Bonnell I. A., 2004, MNRAS, 355, 543
\bibitem[\protect\citeauthoryear{Rice et al.}{2003}]{Rice2003} Rice W. K. M., Armitage P. J., Bate M. R., Bonnell I. A., 2003, MNRAS, 338, 227
\bibitem[\protect\citeauthoryear{Richard et al.}{2013}]{Richard2013} Richard S., Barge P., Le Dizes S., 2013, A\&A, 559, 30
\bibitem[\protect\citeauthoryear{Shen et al.}{2006}]{Shen2006} Shen Y., Stone J. M., Gardiner T. A., 2006, ApJ, 653, 513
\bibitem[\protect\citeauthoryear{Umurhan \& Regev}{2004}]{Umurhan2004} Umurhan O. M., Regev O., 2004, A\&A, 427, 855
\bibitem[\protect\citeauthoryear{Youdin \& Johansen}{2007}]{Youdin2007} Youdin A., Johansen A., 2007, ApJ, 662, 613
\bibitem[\protect\citeauthoryear{Youdin \& Shu}{2002}]{Youdin2002} Youdin A., Shu F., 2002, ApJ, 580, 494
\bibitem[\protect\citeauthoryear{Zhu et al.}{2014}]{Zhu2014} Zhu Z., Stone J. M., Rafikov R. R., Bai X.-N., 2014, ApJ, 785, 122
\bibitem[\protect\citeauthoryear{Zhu et al.}{2010}]{Zhu2010} Zhu Z., Hartmann L., Gammie C., Book L., Simon J., Engelhard E., 2010, ApJ, 713, 1134
\bibitem[\protect\citeauthoryear{Zhu et al.}{2009}]{Zhu2009} Zhu Z., Hartmann L., Gammie C., 2009, ApJ, 694, 1045

\end{thebibliography}
\end{document}